\documentclass[twocolumn,aps,prl,superscriptaddress,nofootinbib]{revtex4-1}
\usepackage{graphicx}
\usepackage{times}
\usepackage{bm}
\usepackage[linktocpage=true,colorlinks=true,urlcolor=blue,citecolor=red]{hyperref}
\usepackage{pgfplots}
\usepackage{amsfonts}
\usepackage{amsmath}
\usepackage{xcolor}
\usepackage{braket}
\usepackage{relsize}
\usetikzlibrary{arrows}
\usepackage{enumerate}
\usetikzlibrary[patterns] 
\usetikzlibrary[snakes] 
\usetikzlibrary[shapes] 
\usepackage{pgfplots}
\usepackage{newfloat}

\DeclareFloatingEnvironment[name={Supplementary Figure}]{suppfigure}

\graphicspath{ {Figures/} }
\begin{document}

\title{Electronic structure of full-shell InAs/Al hybrid semiconductor-superconductor nanowires: Spin-orbit coupling and topological phase space}

\author{Benjamin D. Woods}
\affiliation{Department of Physics and Astronomy, West Virginia University, Morgantown, WV 26506, USA}
\author{Sankar Das Sarma}
\affiliation{Condensed Matter Theory Center and Joint Quantum Institute, Department of Physics, University of Maryland, College Park, Maryland, 20742-4111, USA}
\author{Tudor D. Stanescu}
\affiliation{Department of Physics and Astronomy, West Virginia University, Morgantown, WV 26506, USA}
\affiliation{Condensed Matter Theory Center and Joint Quantum Institute, Department of Physics, University of Maryland, College Park, Maryland, 20742-4111, USA}

\begin{abstract}
We study the electronic structure of full-shell superconductor-semiconductor nanowires, which have recently been proposed for creating Majorana zero modes, using an eight-band $\vec{k} \cdot \vec{p}$ model within a fully self-consistent Schr\"{o}dinger-Poisson scheme.  We find that the spin-orbit coupling induced by the intrinsic radial electric field is generically weak for sub-bands with their minimum near the Fermi energy. Furthermore, we show that the chemical potential windows consistent with the emergence of a topological phase are small and sparse and can only be reached by fine tunning the diameter of the wire. These findings suggest that the parameter space consistent with the realization of a topological phase in full-shell InAs/Al nanowires is, at best, very narrow.
\end{abstract}

\maketitle

Hybrid semiconductor-superconductor (SM-SC) nanowires have recently become the subject of intense research in the context of the quest for topological  Majorana zero modes (MZMs) \cite{Read2000,Kitaev2001}. Motivated by the promise of fault-tolerant topological quantum computation \cite{Nayak2008,DSarma2015} and following concrete theoretical proposals \cite{Sau2010a,Lutchyn2010,Oreg2010}, this nanowire-based MZM search has shown impressive experimental progress in the past few years \cite{Mourik2012,Deng2012,Das2012,Churchill2013,Finck2013,Albrecht2016,Deng2016,Nichele2017,Zhang2017,Gul2018}. 
Nonetheless, reaching the level of the definitive demonstration and consistent realization of isolated MZMs requires further development and  improvement. The lack of definitive evidence of topological Majorana behavior, e.g., correlated tunneling features at the opposite ends of the system \cite{DSarma2012}, and the real possibility of having trivial low-energy Andreev bound states (ABSs) mimicking the MZM phenomenology \cite{Kells2012,Moore2018,Liu2017a,Vuik2019,Reeg2018b,Stanescu2018b}, instead of actual MZMs, underscore the importance of being able to finely control the electrochemical potential in gated devices and to engineer structures with large effective g-factors and spin-orbit couplings, which represent    
key necessary conditions for creating/stabilizing nanowire-based MZMs.

To alleviate some of these rather stringent requirements and the associated problems, an alternative path to creating MZMs, which uses magnetic flux applied to SM wires coated with a full SC shell, was recently proposed \cite{Vaitiekenas2018,Lutchyn2018b}. This scheme eliminates the need for a large Zeeman splitting (i.e. large effective g-factor or large magnetic field) and also generates a more uniform and reproducible electrostatic environment (which may help avoid creating trivial ABSs). The main disadvantages of this approach are the impossibility of directly controlling the chemical potential using gates and absence of a large electric field across the wire to ensure strong spin-orbit coupling. While the chemical potential can be tuned by controlling the diameter of the wire (i.e. using different samples),  a spin-orbit coupling strength on the order of $200~$meV \AA$~$(or larger) is required to access the topological phase \cite{Lutchyn2018b}. Since these parameters cannot be directly measured experimentally, obtaining reliable theoretical estimates represents an essential task. To capture the basic physics, it is critical to take into account i) the multi-orbital nature of the SM bands (by incorporating at least $s$- and $p$-orbital contributions) and ii) the electrostatic effects (by self-consistently solving a Schr\"{o}dinger-Poisson problem). We note that these are crucial issues for the entire research field of SM-SC hybrid nanostructures, but they have only recently started to be addressed, and only within single-orbital approaches \cite{Vuik2016,Woods2018,Mikkelsen2018,Antipov2018}.

In this work, we determine the spin-orbit coupling, chemical potential, and effective mass for full-shell InAs/Al nanowires based on an eight-band $\vec{k} \cdot \vec{p}$ model \cite{Winkler2003} using a mean-field treatment of the long-range electron-electron interaction within a fully self-consistent Schr\"{o}dinger-Poisson scheme. We find that the chemical potential windows consistent with the emergence of a topological phase form a sparse set and require extreme fine tunning of the wire diameter. Furthermore,  we find that the spin-orbit coupling is weak (on the order of $30-60~$ meV \AA) for all physically-relevant values of the wire diameter and SM-SC work function difference, making any emergent topological superconducting phase very weakly protected by a small gap. Based on these findings, we conclude that realizing topological superconductivity and MZMs in full-shell SM-SC nanowires represents a low-success-probability target. If realized, the topological phase is likely to be characterized by a small topological gap. We also provide suggestions for possible optimizations of the full-shell scheme.  

\begin{figure}[t]
\begin{center}
\includegraphics[width=0.48\textwidth]{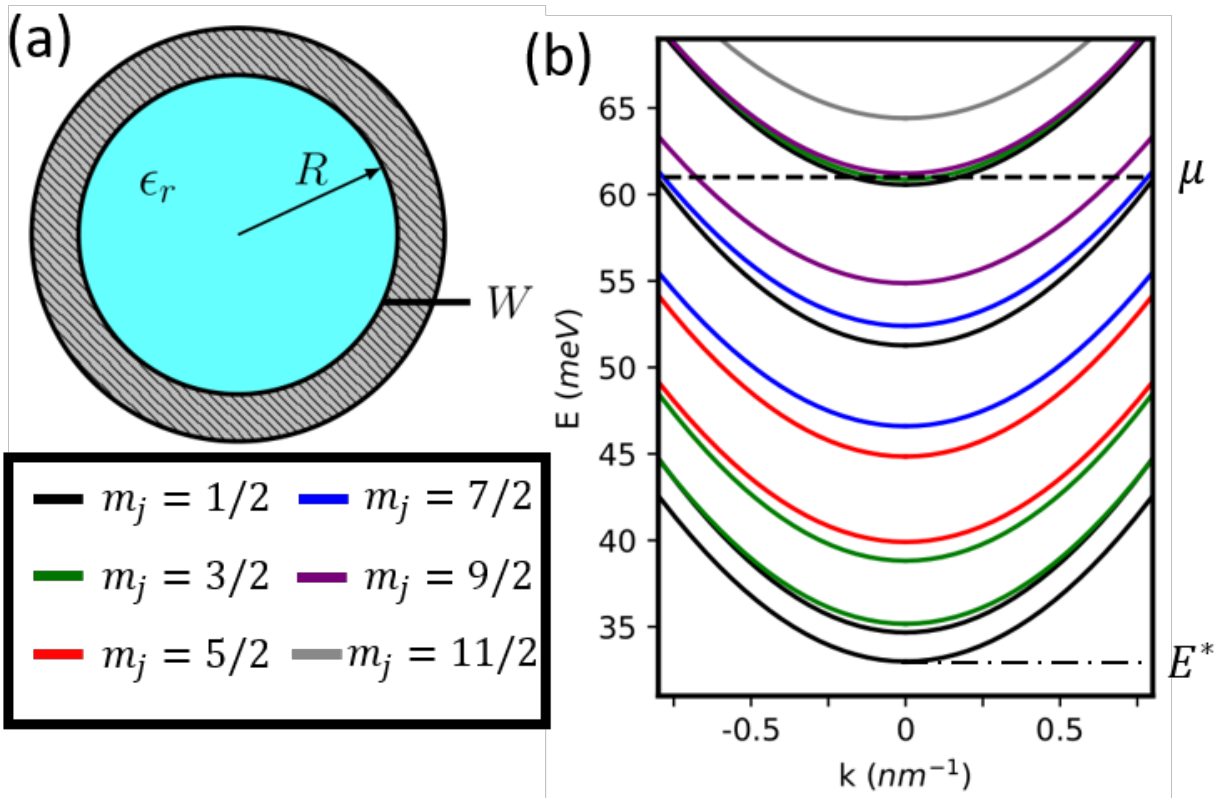}
\end{center}
\vspace{-4mm}
\caption{(a) Cross section of a full-sell nanowire consisting of an InAs core (blue) and an Al shell (gray). The SC-SM interface (characterized by the work function difference $W$) is treated as a Dirichlet boundary condition. (b) Low energy conduction band structure obtained by solving Eqs. (\ref{Ham}-\ref{Poisson}) self-consistently. The colors designate pairs of bands corresponding to given $m_J$ quantum numbers (labeling the z-component of total angular momentum). The chemical potential $\mu$ and zero point energy $E^{*}$ are marked by the dashed and dashed-dotted lines, respectively.}
\label{FIG1}
\vspace{-1mm}
\end{figure}

We consider a cylindrical full-shell nanowire, as represented schematically in Fig. \ref{FIG1}(a). The SM core is modeled using an eight-band $\vec{k} \cdot \vec{p}$ model \cite{Winkler2003,Kohn1955,Luttinger1956} in the presence of a mean-field effective potential, 
\begin{equation}
H = H_{\vec{k} \cdot \vec{p}} - e~\!\phi\left({r}\right),  \label{Ham}  
\end{equation}
where the mean field potential $\phi\left(r\right)$ is induced by the net charge inside the SM wire and must be determined self-consistently.  
While other approaches, such as density functional theory and empirical tight-binding methods \cite{Chang1988,Loehr1994}, are known to accurately capture the electronic structure of semiconductors, $\vec{k} \cdot \vec{p}$ methods are much less computationally demanding and are quite accurate near the high symmetry points of the Brillouin zone, which are of interest here \cite{Marquardt2008}.  Note that  InAs nanowire grown along the [111] crystallographic direction have a hexagonal cross section, but the cylindrical approximation used here for simplicity is expected to be quite	accurate \cite{Luo2016}. In addition, we adopt the so-called axial approximation, which amounts to promoting the underlying atomic fcc lattice symmetry to a full rotation symmetry  abut the z-axis \cite{Winkler2003}, so that the z-component of the angular momentum, $J_z$, is conserved (see the Supplementary Material for details).

The mean field potential $\phi\left(r\right)$ is determined by solving the Poisson equation 
\begin{equation}
\nabla^2 \phi\left(r\right) = -\frac{\rho\left(r \right)}{\epsilon}, \label{Poisson}
\end{equation}
where $\rho$ is the charge density corresponding to the occupied conduction band states and $\epsilon =\epsilon_r \epsilon_o$,  with $\epsilon_r=15$, is the lattice dielectric constant of InAs. The chemical potential is determined by the work function difference between the SM  and the SC ($W$) and by the energy of conduction band edge ($E_o$ for bulk InAs). In the full-shell geometry, $W$ and $E_o$ are not independent parameters (as they are in a ``standard'' gated configuration, where the chemical potential $\mu$ is tuned independently) and they can be combined as
\begin{equation}
\mu = W - E_o.  
\end{equation}
With this definition of the chemical potential, the boundary condition at the SM-SC interface \cite{Woods2018,Antipov2018,Mikkelsen2018} becomes $\phi(R)=0$, as the global band shift due to the work function difference is already incorporated in $\mu$. 
Finally, we note that the SC shell is not explicitly included in our model, but serves as an electrostatic boundary condition (through the work function difference $W$). While the presence of a SC is known to renormalize the band structure of the hybrid system \cite{Cole2015,Stanescu2017a,Liu2017a,Reeg2018a,Woods2018,Antipov2018,Mikkelsen2018}, the goal of this work is to determine the ``bare'', i.e. unrenormalized  wire parameters characterizing the self-consistent electronic structure of the full-shell system. 

The Schr\"{o}dinger equation, $H\Psi=E\Psi$, where $\Psi$ is an eight component spinor, and the Poisson equation (\ref{Poisson}) are solved self-consistently. For the cylindrical geometry and within the axial approximation, we have
\begin{equation}
\Psi_{m_J}\left(\vec{r},k_z\right) = \frac{g(r,k_z)}{\sqrt{r}} e^{ik_zz} e^{i(m_J - M_{s})\varphi}, \label{PSI}
\end{equation}
where $g(r,k_z)$ is an eight component spinor, $m_J \in (\mathbb{Z} + \frac{1}{2})$ labels the z-component of the total angular momentum, and $M_s = {\rm diag} \left(\frac{1}{2},-\frac{1}{2},\frac{3}{2},\frac{1}{2},-\frac{1}{2},-\frac{3}{2},\frac{1}{2},-\frac{1}{2}\right)$ is a diagonal matrix. The first two entries represent s-orbitals, the next four are p-orbitals with angular momentum $j=3/2$, and the last two are p-orbitals with $j=1/2$.  

The band structure for a prototypical full-shell wire of radius $R=45~$nm with $\mu = 62~$meV is shown in Fig. \ref{FIG1}(b). Only the conduction sub-bands are shown. At zero magnetic field, the states corresponding to $m_J$ and $-m_J$ have the same energy, hence all sub-bands are double degenerate. Note that each $m_J$ value corresponds to two sub-bands separated by a finite energy gap. The sub-bands consist of  nearly opposite spin states with dominant orbital angular momentum $\ell$ and $\ell+1$. 
All states up to the chemical potential $\mu$ (dashed line in Fig. \ref{FIG1}) are filled. The energy  $E^*$ corresponding to the bottom of the conduction band is the zero point energy due to finite size confinement and the mean-field effective potential $\phi(r)$. 

\begin{figure}[t]
\begin{center}
\includegraphics[width=0.48\textwidth]{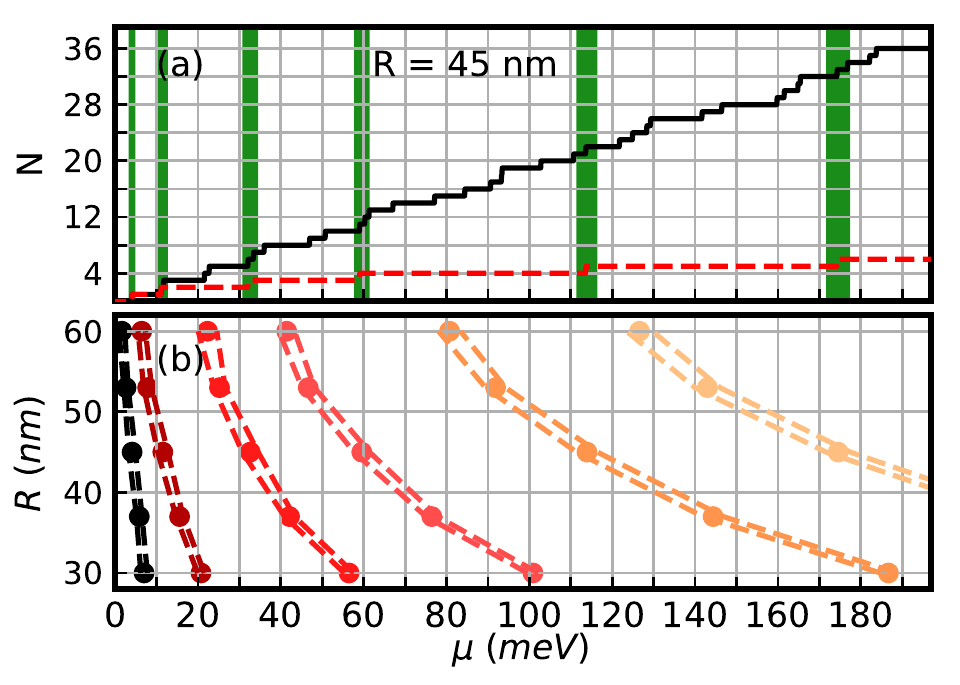}
\end{center}
\vspace{-4mm}
\caption{(a) Number of occupied $m_J \geq \frac{1}{2}$ and $m_J = \frac{1}{2}$ subbands as a function of $\mu$ are shown in black (solid) and red (dashed) lines, respectively, for a wire of radius $R=45$ $nm$. Green shaded regions show when the bottom of an $m_J=\frac{1}{2}$ subband is within $0.5$ $meV$ of the chemical potential. (b) Dashed lines show boundaries of when $m_J=\frac{1}{2}$ subband is within $0.5$ $meV$ of the chemical potential as function of $\mu$ and $R$. Note that for any $\mu$ there is a subband crossing for a suitable radius $R$. }
\label{FIG2}
\vspace{-1mm}
\end{figure}

The emergence of a topological SC phase supporting  MZMs in cylindrical full-shell nanowires requires a finite magnetic field  inducing a phase winding in the superconducting order parameter and a chemical potential lying near the bottom of an $m_J = \frac{1}{2}$ sub-band \cite{Lutchyn2018b}. 
To determine the likelihood of the chemical potential satisfying this condition, we calculate the band structure of a nanowire of radius $R=45~$nm as a function of $\mu$, i.e. the work-function difference $W$. The (total) number of occupied $m_J \geq \frac{1}{2}$ sub-bands, as well as the number of $m_J = \frac{1}{2}$ sub-bands, are shown in Fig. \ref{FIG2}(a). While $W$ and $E_o$ (hence $\mu$) are not  precisely known, one would expect a chemical potential on the order $\sim10^2~$meV. As shown in Fig. \ref{FIG2}(a), this corresponds to a large number of occupied sub-bands (tens of bands). In addition, the system has a few occupied $m_J = \frac{1}{2}$ sub-bands (red dashed line). The values of $\mu$ consistent with the chemical potential being within $\pm 0.5~$meV of the bottom of an $m_J=\frac{1}{2}$ sub-band (i.e. within an energy window about four times the induced gap) are marked by the green shadings. 
These  regions correspond to (rather optimistic estimates of) parameter values consistent with the emergence of MZMs \cite{Lutchyn2018b}. Note that the width of these regions increases with $\mu$, because the mean field potential increases and it becomes more ``expensive'' to add charge to the system. At the same time, however, the ``green regions'' become more sparse. Basically, Fig. \ref{FIG2}(a)  demonstrates that, for a full-shell wire of radius  $R=45~$nm, the likelihood of satisfying conditions (i.e. having $W$ and $E_o$ values) consistent with the emergence of MZMs is rather low. 
To establish the dependence of this likelihood on the wire radius, we perform self-consistent band structure calculations for different values of $R$ and identify the regions of ``suitable'' chemical potential. The results are shown  in Fig. \ref{FIG2}(b). Note that the intervals between the ``suitable'' regions decrease with increasing radius. Also, since $\mu$ should be independent of $R$ (as it is determined by the SC-SM work function difference $W$),  Fig. \ref{FIG2}(b) shows that the system can be brought into a regime consistent with the emergence of MZMs by (finely) tunning the radius of the wire within the $30-60~$nm range. Note, however, that the fine tunning requirement becomes more stringent at large values of the chemical potential. This also implies that, if a wire of radius $R$ supports a topological SC phase, wires with slightly different radii, e.g. $R\pm 5~$nm, should {\em not} be able to support topological phases. Finally, we emphasize that these considerations hold under the assumption that the value of the work function difference, $W$, is relatively stable from device to device (otherwise, the realization of the topological condition becomes purely a matter of chance and wild luck).  

Next, we investigate the spin-orbit coupling and extract effective parameters for the 2-band  model Hamiltonian $H_{eff}$ recently used to study the topological properties of full-shell nanowires \cite{Lutchyn2018b}. Explicitly, we have
\begin{equation}
H_{eff} = \frac{\hbar^2 k^2}{2m^*} - \mu + \alpha \hat{r} \cdot
\left[\vec{\sigma} \times \vec{k}\right],   \label{Heff}
\end{equation}
where $m^*$ is the effective mass, $\mu$ is the chemical potential, $\sigma_i$ ($i=x,y,z$) are the Pauli spin matrices, and $\alpha$ is a phenomenological spin-orbit coupling  coefficient. Again, since the system has cylindrical symmetry, $m_J \in (\mathbb{Z} + \frac{1}{2})$ is a good quantum number and each $m_J$ value labels a pair of sub-bands separated by a $k_z$-dependent energy gap. We determine the spin-orbit coupling $\alpha$ and the effective mass $m^*$ by fitting a given pair of sub-bands of the full 8-band model with the corresponding $m_J$ pair of the effective Hamiltonian (\ref{Heff}). The details of the extraction procedure are provided in the Supplementary Material. 

\begin{figure}[t]
\begin{center}
\includegraphics[width=0.48\textwidth]{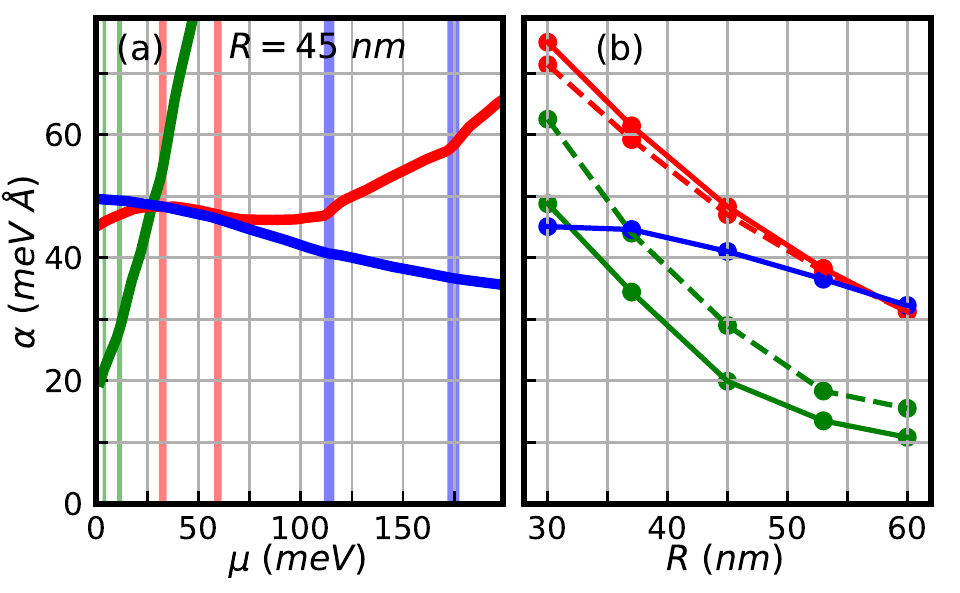}
\end{center}
\vspace{-4mm}
\caption{(a) Spin-orbit coupling coefficient $\alpha$ as function of $\mu$ for the first three $m_J = \frac{1}{2}$ sub-band pairs (green, red, and blue lines, respectively) in a wire with $R=45~$nm. The sub-band bottoms lie within  $\pm 0.5~$meV of the chemical potential within the corresponding shaded regions [see also Fig \ref{FIG2}(a)].
 (b) Spin orbit coupling coefficient $\alpha$ for the $m_J = \frac{1}{2}$ sub-band pair at chemical potential crossings as function of $R$. The colors correspond to those in panel (a) with the solid and dashed lines denoting the first and second crossing of a sub-band pair, respectively.}
\label{FIG3}
\vspace{-1mm}
\end{figure}

The effective spin-orbit coupling coefficients corresponding to the $m_J=\frac{1}{2}$ states for a wire of radius $R=45~$nm are shown in Fig. \ref{FIG3}. Only the first three pairs are represented, as the higher energy pairs occur for $\mu > 200~$meV, but we checked that the main features hold for larger values of the chemical potential. The spin-orbit coefficient associated with the first $m_J=\frac{1}{2}$ pair (shown in green) increases nearly linearly with $\mu$, i.e., with the work function difference $W$ [see panel (a)]. However, this pair is relevant for topological physics only in the regime $\mu < 10~$meV, when it is close-enough to the chemical potential (shaded green ranges). Similarly, the relevant values of $\alpha$ associated with the higher energy pairs are those within the corresponding ``topological'' windows, as determined in Fig. \ref{FIG2}(a). For example, if $\mu\approx 175~$meV, the only relevant contribution to a possible topological phase is given by the second component of the third $m_J=\frac{1}{2}$ pair, which is characterized by $\alpha\approx 37~$meV \AA. Although the second pair has $\alpha\approx 58~$meV \AA$~$ and the first pair has an even larger spin-orbit coupling, they are very far from the chemical potential and cannot induce a TQPT. 

The main result shown in Fig. \ref{FIG3}(a) is that the effective spin-orbit coupling of $m_J=\frac{1}{2}$ sub-bands located  in the vicinity of the chemical potential does not exceed $50~$mev \AA$~$ in a wire  of radius $R=45~$nm, {\em regardless of the work function difference}. 
To determine the dependence of the spin-orbit coupling strength on the radius of the wire, we calculate the effective coupling of $m_J=\frac{1}{2}$ sub-bands that lie in the vicinity of the chemical potential for wires with $30\leq R \leq 60~$nm. The results are shown in Fig. \ref{FIG3}(b). First, we note that for a given $m_J=\frac{1}{2}$ pair the spin-orbit coupling (at the chemical potential) decreases with increasing wire radius. Qualitatively, this can be understood as follows: increasing $R$ reduces the inter-band spacings, so that the chemical potential crossing (for a given sub-band) will occur at a lower value of $\mu$, i.e. in the presence of less charge inside the wire, hence a weaker mean-field potential. In turn, the reduced potential generates a weaker spin-orbit coupling. The second property revealed by the results shown in panel (b) is that  the overall magnitude  of the spin-orbit coupling for $m_J=\frac{1}{2}$ sub-bands in the vicinity of the chemical potential remains small (i.e. $\alpha < 75~$meV\AA) {\em regardless of radius}, i.e. for wires with $30\leq R \leq 60~$nm and arbitrary work function (so that $0<\mu<200~$meV). We remind the reader that the predicted spin-orbit coupling strength required for the realization of topological superconductivity is on the order of $200~ $meV\AA$~$ (or larger) \cite{Lutchyn2018b}. The central result of this work, shown in Fig. \ref{FIG3}, demonstrates that such values of the effective spin-orbit coupling cannot be realized in full-shell InAs nanowires. Note that reducing the radius of the wire may increase the effective spin-orbit coupling, but finding a radius that is consistent with the emergence of topological superconductivity may become a challenging task, as discussed in the context of Fig. \ref{FIG2}(b). The whole procedure then becomes a matter of time-consuming trial and error dependent on getting `lucky'.



\begin{figure}[t]
\begin{center}
\includegraphics[width=0.48\textwidth]{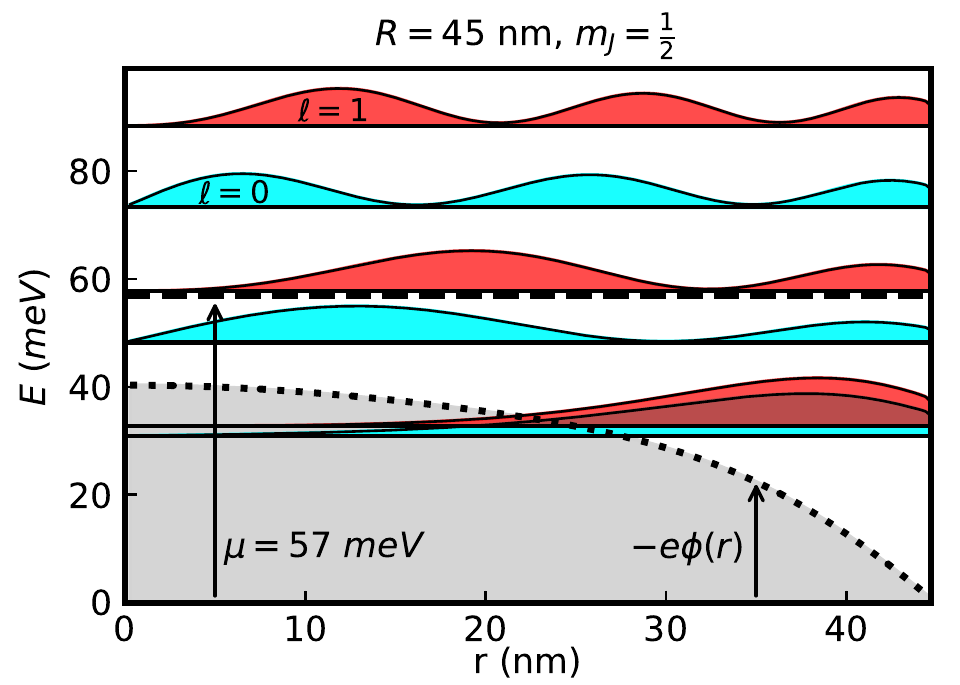}
\end{center}
\vspace{-4mm}
\caption{Wave function profiles, $\left|\psi\right|^{2}$, of the first six $m_J = \frac{1}{2}$ states at $k_z = 0$ for a wire of radius $R=45~$nm and $\mu = 57~$meV. The states are shifted vertically by their energies. The effective mean-field potential is also show as a dotted line (gray filling), while the chemical potential is marked by the black dashed line.  
The  $m_J = \frac{1}{2}$ states dominated by $\ell = 0$ and $\ell = 1$ components are shaded blue and red, respectively.  Notice that the first two states are confined within the outer region  $30\leq r\leq 45~$nm where the radial electric field is maximum, while the other states are distributed over the entire cross section of the wire.}
\label{FIG5}
\vspace{-2.5mm}
\end{figure}

To better understand the physical reason behind the small spin-orbit coupling values at the chemical potential, we calculate the wave functions of the first six $m_J = \frac{1}{2}$ states at $k_z = 0$ for a wire of radius $R=45~$nm with $\mu=57~$meV. The results are shown in Fig. \ref{FIG5}. Note that the wave function amplitudes are shifted with respect to the bottom of the mean field potential (gray shading) by the energies of the corresponding states,  allowing us to visualize the effect of $\phi(r)$ on various states.  The first two states ($p=1$) are localized near the surface of the SM wire (i.e. the SM-SC interface). This is not surprising, as their energy is below the top of the mean-field potential,  which effectively pushes them away from the center of the wire. Since the electric field ${\bm E}=-{\bm \nabla}\phi$ is maximum in the outer region $30\leq r\leq 45~$nm, one would expect a relatively strong spin-orbit coupling for this pair of states ($\alpha > 80~$meV\AA$~$, see Fig. \ref{FIG3}).  
By contrast, the second and third pairs of states have energies well above the potential maximum and are weakly affected by $\phi(r)$. As a result, these states are extended throughout the entire cross section of the wire and the effect of the radial electric field will be strongly suppressed, resulting in lower values of the spin-orbit coupling ($\alpha\approx 47~$meV\AA$~$ in Fig. \ref{FIG3}).  

In conclusion, we studied the electronic structure of full-shell InAs/Al hybrid nanowires using an eight-band $\vec{k} \cdot \vec{p}$ model which was solved within a fully self-consistent  Schr\"{o}dinger-Poisson scheme. We found that the spin-orbit coupling  of the $m_J = \pm\frac{1}{2}$ sub-bands near the chemical potential is generically small $\alpha < 70~$meV\AA, regardless of the chemical potential (i.e. the work functions difference between the SM wire and the SC shell) or the wire diameter. In addition, we demonstrated that bringing the bottom of an $m_J = \pm\frac{1}{2}$ sub-band close to the chemical potential  requires fine tunning the wire radius. More specifically, within the range $30 \leq R \leq 60~$nm one should expect to find about two small windows (each a few nanometers wide) consistent with the presence of an $m_J = \pm\frac{1}{2}$ sub-band near the chemical potential. 
Since the existence of low-energy $m_J = \pm\frac{1}{2}$ sub-bands with strong effective spin orbit coupling is critical for the emergence of a topological phase in full-shell nanowires, our findings suggest that the parameter space consistent with such a phase may be, at best, very narrow.   
As a possible solution for enhancing the spin-orbit coupling, we suggest using core-shell SM wires, with a wide gap material (e.g., GaAs) for the core and a narrow-gap SM (e.g. InAs) for the shell. In essence, the presence of the core will push the states toward the outer region, where the radial electric field is large, increasing the spin-orbit coupling. Finally, we note that the presence of symmetry breaking perturbations (e.g., due to the hexagonal wire geometry) is unlikely to generate a dramatic increase of the spin-orbit coupling and will not change our findings regarding the requirement to fine tune the wire radius. We conclude, therefore, that finding topological Majorana modes in full-shell nanowires will be quite challenging and will depend on considerable trial and error to achieve a lucky sweet spot in optimizing the spin-orbit coupling and chemical potential.  The lack of a suitable tuning parameter in situ is a serious problem in this respect.

\begin{acknowledgments}
This work is supported by Microsoft Q, Laboratory for Physical Sciences, and NSF DMR-1414683.
\end{acknowledgments}

%

\pagebreak
\widetext
\begin{center}
\textbf{\large Supplementary Material: Electronic structure of full-shell InAs/Al hybrid semiconductor-superconductor nanowires: Spin-orbit coupling and topological phase space}
\end{center}
\setcounter{equation}{0}
\setcounter{figure}{0}
\setcounter{table}{0}
\setcounter{page}{1}
\makeatletter
\renewcommand{\theequation}{S\arabic{equation}}
\renewcommand{\thefigure}{S\arabic{figure}}
\renewcommand{\bibnumfmt}[1]{[S#1]}
\renewcommand{\citenumfont}[1]{S#1}

\section{Details of eight-band model} \label{Sup1}
In this section we provide a more detailed account of the eight-band $\vec{k} \cdot \vec{p}$ Hamiltonian used in the main text. Within this eight band model, we write the solution of the Schr\"{o}dinger equation in the form
\begin{equation}
\psi\left(\vec{r}\right) = \sum\limits_\alpha \psi_\alpha\left(\vec{r}\right) \left | u_\alpha \right>,   
\end{equation}
where $\left| u_\alpha \right>$ are Kane basis functions \cite{Kane1957} given by 
\begin{align}
\begin{split}
\left| u_1 \right> = \left| \frac{1}{2},\frac{1}{2}\right>_{s},&\, \left| u_2 \right> = \left| \frac{1}{2},-\frac{1}{2}\right>_{s},\\
\left| u_3 \right> = \left| \frac{3}{2},\frac{3}{2}\right>_{p},&\, \left| u_4 \right> = \left| \frac{3}{2},\frac{1}{2}\right>_{p}, \\
\left| u_5 \right> = \left| \frac{3}{2},-\frac{1}{2}\right>_{p},&\, \left| u_6 \right> = \left| \frac{3}{2},-\frac{3}{2}\right>_{p}, \\
\left| u_7 \right> = \left| \frac{1}{2},\frac{1}{2}\right>_{p},&\, \left| u_8 \right> = \left| \frac{1}{2},-\frac{1}{2}\right>_{p}.
\end{split}
\end{align}
These basis functions $\left|s,M_s\right>_i$ are labeled by total ``spin'' angular momentum, $s$, and ``spin'' angular momentum about the z-axis, $M_s$, respectively. Finally, the subscripts indicate the orbitals from which the basis states derive, with $s$ and $p$ indicating $s$ and $p$ symmetric orbitals, respectively. The total angular momentum, $\vec{J}$, is of course a vector sum between the spin, $\vec{S}$, and orbital, $\vec{L}$ components, respectively. The total angular momentum is generically not conserved within the eight-band Hamiltonian due to the underlying fcc lattice symmetry. However, one can restore the conservation of angular momentum about a given axis through use of the axial approximation \cite{Winkler2003A}. This essentially constrains the Hamiltonian such that $\left[ J_z,H\right]_- = 0$, where $J_z$ is the total angular momentum operator about the z-axis. In our case we wish choose the z-axis to lie along the $\left[111\right]$ crystallographic direction, e.g., the growth direction of the nanowire. The eight band Hamiltonian in the absence of an applied magnetic field and superconductivity is then given by \cite{Winkler2003A,Luo2016A,Ivashev2016}     
\begin{equation}
H_{\vec{k} \cdot \vec{p}} = 
\begin{bmatrix}
A& 0 & V^{\dagger}& 0& \sqrt{3}V& -\sqrt{2}U& -U& \sqrt{2}V^{\dagger}& \\ 
0& A& -\sqrt{2}U& \sqrt{3}V^{\dagger}& 0& -V& \sqrt{2}V& U& \\ 
V& -\sqrt{2}U& -P+Q& -S& R& 0& \frac{1}{\sqrt{2}}S& -\sqrt{2}R& \\ 
0& -\sqrt{3}V& -S^\dagger& -P-Q&0& R& \sqrt{2}Q& \sqrt{\frac{3}{2}}S& \\ 
\sqrt{3}V^\dagger& 0& R^\dagger& 0& -P-Q& S& -\sqrt{\frac{3}{2}}S^\dagger& -\sqrt{2}Q&\\ 
-\sqrt{2}U& -V^\dagger& 0& R^\dagger& S^\dagger& -P+Q& \sqrt{2}R^\dagger& \frac{1}{\sqrt{2}}S^\dagger& \\ 
-U& \sqrt{2}V^\dagger& \frac{1}{\sqrt{2}}S^\dagger& \sqrt{2}Q& -\sqrt{\frac{3}{2}}S& \sqrt{2}R& Z& 0& \\ 
\sqrt{2}V& U& -\sqrt{2}R^\dagger& \sqrt{\frac{3}{2}}S^\dagger& -\sqrt{2}Q& \frac{1}{\sqrt{2}}S& 0& Z& 
\end{bmatrix}, \label{Hkp}
\end{equation}
where
\begingroup
\allowdisplaybreaks
\begin{align}
A &= \frac{\hbar^{2}k^2}{2m_o}, \\
P &= \gamma_1 \frac{\hbar^{2}k^2}{2m_o} - E_v, \\
Q &= -\gamma_3 \frac{\hbar^{2}}{2m_o}\left(k_x^{2} + k_y^{2} - 2k_z^2\right), \\
S &= \frac{-2}{\sqrt{3}}\left(2\gamma_2+\gamma_3\right)k_zk_-, \\
R &= \frac{1}{\sqrt{3}} \frac{\hbar^{2}}{2m_o} \left(\gamma_2 + 2\gamma_3\right) k_-^{2},\\ 
U &= \frac{E_p}{\sqrt{3}} k_z, \\ 
V &= -\frac{E_p}{\sqrt{3}} k_-, \\
Z &= -P - \Delta,
\end{align}
\endgroup
where $m_o$ is the mass of an electron, $\gamma_1$, $\gamma_2$, and $\gamma_3$ are the Luttinger parameters, $\Delta$ is the spin split-off energy, $E_p$ is the s-p coupling strength, and $k_{\pm} = k_x \pm i k_y$. Note that $k_x, k_y \rightarrow -i \frac{\partial}{\partial x}, -i \frac{\partial}{\partial y}$ due to the finite cross section of the wire. Within the axial approximation, $k_{\pm}$ takes a special role as they act as raising and lowering operators of orbital angular momentum about the z-axis. As an example, consider the $S$ operator and where it occurs in (\ref{Hkp}). Note that the $S$ operator contains a $k_-$ operator and couples the $\left|u_3\right>=\left|\frac{3}{2},\frac{3}{2}\right>$ and $\left|u_4\right>=\left|\frac{3}{2},\frac{1}{2}\right>$ basis states. In order to conserve total angular momentum about the z-axis, the z-component of orbital angular momentum must differ by a unit of angular momentum, e.g. $\ell_4 - \ell_3 = 1$, where $l_n$ labels the orbital angular momentum about the z-axis for the $n^{th}$ component. Therefore, $k_-$ is needed to lower the orbital angular momentum of the $\left|u_4\right>$ component and allow coupling between the two basis states. All other operators ($P$,$Q$,$R$, etc.) must also obey this conversation condition. Finally, note that the cylindrical symmetry of the nanowire is also necessary to conserve the z-component of angular momentum.

The conservation of total angular momentum about the z-axis suggests that we write the wavefunctions in the form
\begin{equation}
\psi_{m_J}(\vec{r},k_z) = \sum\limits_\alpha \frac{g_{m_J,\alpha}\left(r,k_z\right)}{\sqrt{r}} e^{i \, \left( \ell_{m_J,\alpha} \varphi + k_z z\right)}, \label{PSI2}  
\end{equation}
where, $g_{m_J,\alpha}$ is the $\alpha$ band component of the eight-component spinor $g_{m_J}$, and $m_J$ and $k_z$ are quantum numbers labeling z-component of total angular momentum and momentum, respectively. The orbital angular momentum of each component is given by $\ell_{m_J,\alpha}$. To satisfy the angular momentum conservation law, $\ell_{m_J,\alpha}$ must be
\begin{equation}
\ell_{m_J,\alpha} = m_J - M_{s,\alpha},   
\end{equation}
where $M_{s,\alpha}$ is the spin angular momentum about the z-axis of the $\alpha^{th}$ basis state. We note that the wavefunctions must be single valued, which implies $m_J \in \mathbb{Z} + \frac{1}{2}$. As an example, $\ell_{m_J,\alpha}$ for $m_J = \frac{1}{2}$ is given by $\ell_{m_J = \frac{1}{2},\alpha} = \left(0,1,-1,0,1,2,0,1\right)$. We stress that the value of $\ell_\alpha$ for a given basis state can have a dramatic impact of the resulting wavefunctions. This is due to the effective potential caused by the angular momentum when the problem is reduced to a 1D radial equation. In particular, the effective potential of $\ell = 0$ basis states is attractive near the origin, while the effective potential all $\ell \neq 0$ is repulsive near the origin. This difference can clearly be seen in the wavefunctions of Fig. \ref{FIG5} of the main text, which are dominated by $\ell = 0$ and $\ell = 1$ basis states, respectively.

Plugging in (\ref{PSI2}) into the Schr\"{o}dinger equation results in a radial matrix equation for $g_{m_J}\left(r\right)$. Note that a separate equation must be solved for each $m_J$ and $k_z$ pair. We solve for $g_{m_J}(r)$ using finite difference method on a uniform radial lattice. We employ standard finite difference approximations for the derivatives given by
\begin{align}
\left.\frac{dg}{dr}\right|_{r_n} &\approx \frac{g(r_{n+1}) - g(r_{n-1})}{2a}, \\
\left.\frac{d^2g}{dr^2}\right|_{r_n} &\approx
\frac{g(r_{n+1}) + g(r_{n-1})- 2g(r_{n})}{a^2},
\end{align}
where $a$ is the lattice spacing. In eq. (\ref{PSI2}) we divided by $\sqrt{r}$ such that $g(r)$ must vanish at $r=0$ for all components, in contrast to the components of $\psi_{m_J,k_z}$, which don't necessarily vanish at $r=0$ \cite{Arsoski2015,Winkler2017}. While simplifying the boundary conditions, discretized radial equations for $g(r)$ are known to suffer from convergence issues due to the a singularity of the effective potential for solutions with basis states of zero orbital angular momentum \cite{Laliena2018,Arsoski2015}. Ref. \cite{Laliena2018} showed how one can fix this convergence issue for single band cases by altering the effective potential using information of the asymptotic solution as $r\rightarrow0$. One can show that all eight components of the $\vec{k} \cdot \vec{p}$ Hamiltonian decouple asympotically as $r\rightarrow0$. Therefore we fix the convergence issues by altering all $r^{-2}$ components of the effective potential in a similar fashion to the single band case described in Ref. \cite{Laliena2018}.  

\section{Solution to the Poisson Equation and Self-Consistency}
To solve a Schr\"{o}dinger-Poisson system, we of course need to be able to solve the Poisson equation. Recall that we incorporate the work function difference, $W$, into the definition of the chemical potention, $\mu$ (see main text), and therefore, we need to solve the Poisson equation with homogeneous boundary conditions at the SM-SC interface. Due to the cylindrical symmetry, we know that the charge density will also have cylindrical symmetric. Moreover, we also assume that the charge density is uniform within the unit cell of each grid point of the radial lattice. Therefore, the charge density will be made up of a linear combination of cylindrical shells centered at the various radial positions of the radial lattice. We can therefore solve the Poisson equation by solving for the Green's function of each of these cylindrical shells.

\begin{figure}[t]
\begin{center}
\includegraphics[width=0.38\textwidth]{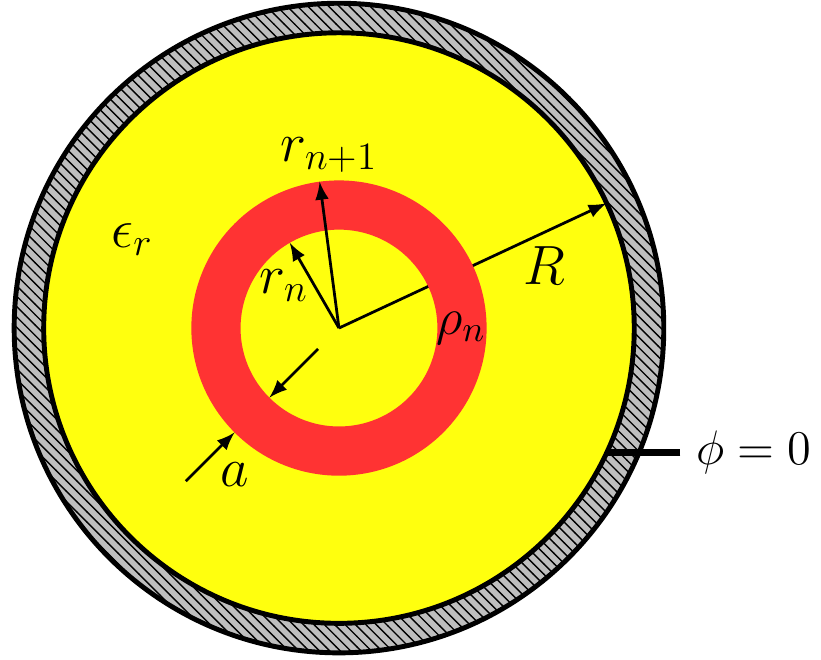}
\end{center}
\vspace{-.4cm}
\caption{Schematic of electrostatic problem to find the Green's function $G_n$. The wire has a radius $R$ with dielectric constant $\epsilon_r$. The SM-SC is set to zero potential, i.e. $\phi\left(r = R\right) = 0$. The $n^{th}$ unit cell is shown in red, with its inner and outer radius given by $r_n$ and $r_{n+1}$, respectively. The charge density $\rho_n$ is assumed to be constant within the unit cell (red area).}
\label{FIGS1}
\vspace{-1mm}
\end{figure}

A schematic of one of these cylindrical shells is shown in Fig. (\ref{FIGS1}). This particular electrostatics problem has an analytic solution that can be found through a simple application of Guass' law. The solution $G_n$ is given by
\begin{equation}
G_n(r) = \left(\frac{\lambda_n}{2\pi\epsilon}\right)
\begin{cases} 
      \frac{1}{2} + \ln\left(\frac{R}{r_{n+1}}\right) -
      \frac{r_n^{2}}{r_{n+1}^{2} - r_n^{2}} \ln\left(\frac{r_{n+1}}{r_n}\right),
      & r < r_n \\
      \frac{1}{2}\frac{r_{n+1}^2 - r^2}{r_{n+1}^2 - r_n^2} + \ln\left(\frac{R}{r_{n+1}}\right) -
      \frac{r_n^{2}}{r_{n+1}^{2} - r_n^{2}} \ln\left(\frac{r_{n+1}}{r}\right), & r_n\leq r\leq r_{n+1} \\
      \ln\left(\frac{R}{r}\right), & r > r_{n+1} 
   \end{cases}
\end{equation}
where $\lambda_n$ and $\epsilon$ are the linear charge density in the $n^{th}$ unit cell and dielectric constant, respectively. To determine the average electrostatic potential $\phi_{nn^{\prime}}$ that an electron feels in the $n^{\prime}$ radial unit cell due the charge in the $n$ unit cell, we average $G_n\left(r\right)$ over the $n^{\prime}$ unit cell. We find
\begin{equation}
\phi_{nn^{\prime}} = \left(\frac{\lambda_n}{2\pi\epsilon}\right)
\begin{cases}
\frac{1}{2} + \ln\left(\frac{R}{r_{n+1}}\right) - \frac{ r_n^{2}}{r_{n+1}^{2} - r_n^{2}}\ln\left(1 + \frac{a}{r_n}\right), & n^\prime < n \\
\frac{1}{2} \frac{r_{n+1}^2 - \left<r^2\right>_n}{r_{n+1}^2 - r_n^2} +
\ln\left(\frac{R}{r_{n+1}}\right) - \frac{ r_n^{2}}{r_{n+1}^{2} - r_n^{2}}
\left[\ln\left(r_{n+1}\right) - \left<\ln r\right>_n\right], & n^\prime = n \\
\frac{1}{2} + \left(\frac{1}{r_{n+1}^2 - r_n^2}\right)
\left[r_{n+1}^2 \ln\left(\frac{R}{r_{n+1}}\right) - r_n^2 \ln\left(\frac{R}{r_n}\right)\right], & n^\prime > n \\
\end{cases}
\end{equation}
where
\begin{align}
\left<r^2\right>_n &= \frac{1}{2} \frac{r_{n+1}^4 - r_n^4}{r_{n+1}^2 - r_n^2}, \\
\left<\ln r\right>_n &= \frac{1}{2} \frac{1}{r_{n+1}^2 - r_n^2} 
\left[ r_{n+1}^2 \left(2\ln r_{n+1} - 1\right) - r_n^2 \left(2\ln r_n -1\right)\right],
\end{align}
and $a$ is the length of a radial unit cell. Finally, we obtain the potential felt in the $n^\prime$ unit cell, $V_{n^\prime}$, by summing over the contributions from all of the unit cells. Explicity,
\begin{equation}
V_{n^\prime} = -e\sum \limits_n \phi_{nn^\prime},   \label{V} 
\end{equation}
where $e$ is the elementary charge. Lastly, we note that $\rho$ is the free charge density within the wire. The charge density is found by summing over the occupied states,
\begin{equation}
\rho\left(\vec{r}\right) 
= \sum_{i}^\bullet \sum_\alpha
\int \frac{dk_z}{2\pi} \left|\psi_{i\alpha}\left(\vec{r},k_z\right)\right|^2 f(E_i(k_z)),   
\end{equation}
where $\psi_{i\alpha}(k_z)$ is the $\alpha$ basis state component of the $i$ eigenstate with wavenumber $k_z$, $f(E) = \left( e^{(E-\mu) / k_B T} + 1\right)^{-1}$ is the Fermi function, $E_i$ is the energy of the $i$ eigenstate, and the $\bullet$ indicates that we only sum over the conduction sub-bands. We can use this expression and the assumption that the charge density is uniformly distributed throughout a unit cell to find $\lambda_n$
in terms of the spinor $g_{m_J}$ defined in Eq. (\ref{PSI2}),
\begin{equation}
\lambda_n = -e\sum_{i}^\bullet \sum_{m_J} \sum_{\alpha} 
\int \frac{dk_z}{2\pi} 
\left|g_{i,m_J,\alpha} \left(r_n,k_z\right) \right|^2
f\left(E_{i,m_J}(k_z)\right), \label{lambdaCalc}
\end{equation}
where $g_{i,m_J,\alpha}$ is the $\alpha$ band component of the $i$ eigenstate with angular momentum quantum number $m_J$. In practice we diagonalize the Hamiltonian for a finite number of $k_z$ values and interpolate the solution to perform the integral in Eq. (\ref{lambdaCalc}).

The total Hamiltonian is then given by
\begin{equation}
H = H_{\vec{k} \cdot \vec{p}} + V.   \label{Ham} 
\end{equation}
The potential $V$ must be solved for self-consistently, meaning the potential that is input into the Hamiltonian (\ref{Ham}) must agree with the potential calculated by Eq. (\ref{V}) using the eigenstates of the Hamiltonian. We use a simple iterative mixing scheme to solve the system self-consistently. To help with convergence we use a small temperature $k_B T = 0.01~$meV. We iterate until the average error of the potential $V_n$ is less than $0.03~$meV. Explicity the error is given by 
\begin{equation}
\left<\Delta V\right> = \frac{1}{N} \sum_{n=1}^{N} \left| V_{n}^{input} - V_{n}^{output}\right|,
\end{equation}
where $N$ is the number of lattice sites.

\section{Extraction of Spin-orbit Coefficient and Effective Mass}
\begin{figure}[t]
\begin{center}
\includegraphics[width=0.6\textwidth]{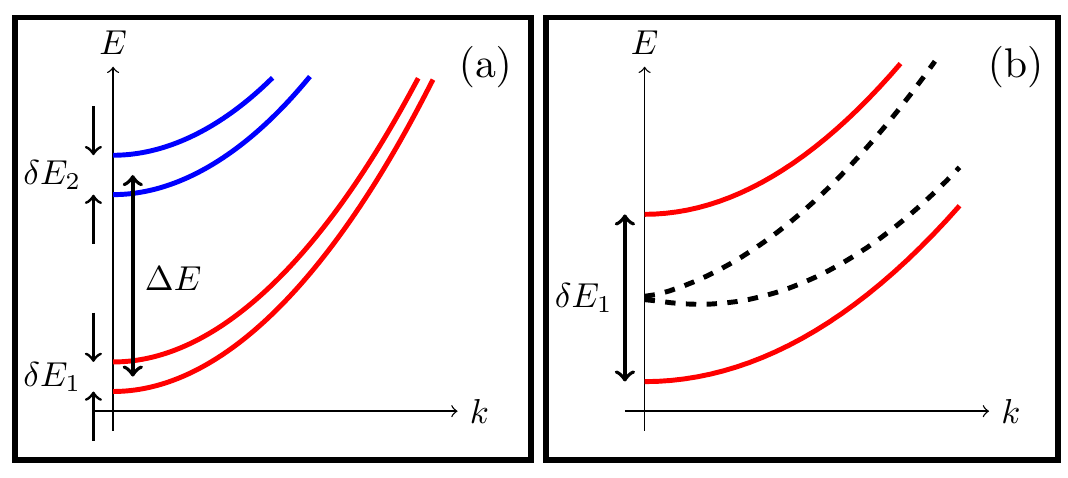}
\end{center}
\vspace{-.4cm}
\caption{(a) Schematic of single $m_J$ sector band structure for either the two-band or eight-band model. Note that only the first four (conduction) sub-bands are shown, and we only show $k \geq 0$. The sub-bands come in pairs as indicated by the coloring of the lines, with the energy spacing between these pairs being much larger than the intra-pair spacing, i.e. $\Delta E \gg \delta E_1, \delta E_2$. The pairs approximately behave as decoupled two-state systems. (b) Red (solid) lines show zoomed in look at sub-bands from the first sub-band pair of panel (a). The effect of spin-orbit interaction is suppressed by the energy spacing $\delta E_1$ between the sub-bands. Black (dashed) lines show the band structure of the perturbed system ($\widetilde{H}$) in which the two sub-bands are degenerate at $k=0$. The spin-orbit effect now becomes pronounced.}
\label{FIGS2}
\vspace{-1mm}
\end{figure}

In this section we describe how we extract the spin-orbit coefficients, $\alpha$, and effective masses, $m^{*}$, of the various sub-bands from the band structure of the eight-band model. To begin we start with the two-band effective model \cite{Lutchyn2018C}
\begin{equation}
H_{eff} = \frac{\hbar^2 k^2}{2m^*} - \mu + \alpha \hat{r} \cdot
\left[\vec{\sigma} \times \vec{k}\right], \label {HamEff}
\end{equation}
where $m^*$ is the effective mass, $\mu$ is the chemical potential, $\alpha$ is the phenomenological spin-orbit coefficient, and $\sigma_i$ $\left(i=x,y,z\right)$ are the Pauli spin matrices. This Hamiltonian respect cylindrical symmetry, which allows us to label states by the z-component angular momentum quantum number $m_J \in \mathbb{Z} + \frac{1}{2}$, just as we have done in the eight-band case. This implies that states with differing $m_J$ quantum numbers do not mix, and we can focus on a single $m_J$ sector. Its illuminating to inspect the band structure from a single $m_J$ sector as we have done in Fig. \ref{FIGS2}(a). First of all, note that the sub-bands come in pairs as indicated by the line colors. To understand why this occurs, note from Eq. (\ref{HamEff}) that the two bands decouple at $k=0$, and therefore the $k=0$ states are composed of only a single band component. Note that two band components must differ by a unit of orbital angular momentum to conserve the total angular momentum. Therefore the intra-pair energy spacings, $\delta E_1$ and $\delta E_2$, are due to the difference in orbital angular momentum between the two band components. Importantly, these intra-pair spacings are generically much smaller than the inter-pair spacing, i.e. $\Delta E \gg \delta E_1, \delta E_2$. This implies that each sub-band pair approximately behaves as a two-state system with the generic Hamiltonian
\begin{equation}
H_{2}\left(k_z\right) = \left(E_{o}+\frac{\hbar^2 k_z^2}{2m^*}\right)\sigma_o + \frac{\delta E}{2}\sigma_z + \widetilde{\alpha}k_z \sigma_y, \label{Ham2}    
\end{equation}
where $E_o$ is the average $k_z = 0$ energy of the sub-band pair, $\delta E$ is the gap between the two levels at $k_z=0$, and $\widetilde{\alpha} = \alpha \chi_o$, where $\chi_o=\left<g_1(k_z=0)\right|\sigma_x\left|g_2(k_z=0)\right>$ (i.e. the overlap between the two states at $k_z=0$). This results in the spectrum
\begin{equation}
E_\pm(k_z) = \left(E_o + \frac{\hbar^2 k_z^2}{2m^*}\right) \pm \left(\frac{\delta E}{2}\right) \sqrt{1 + \frac{4\widetilde{\alpha}^2 k_z^2}{\delta E^2}}. \label{Epm}  
\end{equation}
For states localized away from $r=0$, one finds $\widetilde{\alpha} \approx \alpha$ since the difference in orbital angular momentum between the two band components doesn't significantly impact the wavefunctions, so $\alpha$ and $m^*$ can in principle be determined by fitting the band structure to this dispersion (\ref{Epm}). 

Applying this fitting procedure to the conduction sub-bands of the eight-band model runs into two issues; (1) the effective masses of the two sub-bands within a given sub-band pair differ due to different interactions with the sub-bands derived from the p-orbitals, and (2) the effect of spin-orbit coupling is often dominated by the energy gap $\delta E$. One can work out the dispersion relation by allowing different $m^*$ for the two sub-bands and attempt to fit that dispersion. However, the resulting $\alpha$ is quite uncertain except for the first sub-band pair due to $\delta E$ dominating over $\alpha$.

To overcome these issues we use the following trick to extract $\alpha$ for the $p^{th}$ subband pair. First we diagonalize the eight-band Hamiltonian at $k_z = 0$. This of course results in a diagonal matrix, i.e. $U^{\dagger} H(k_z=0) U = \Lambda =$ diag$(\lambda_1,\lambda_2,\dots)$, where $\lambda_i$ is the energy of the $i$ state. Next we construct the matrix $\widetilde{\Lambda} = \Lambda - \Xi$, where $\Xi$ is another diagonal matrix that shifts the energies of the sub-bands within the $p$ sub-band pair by $\pm \frac{\delta E}{2}$, such that the two sub-bands have the same energy. Lastly, we transform $\widetilde{\Lambda}$ back into the original basis, $\widetilde{H} = U \widetilde{\Lambda} U^{\dagger}$, and diagonalize this perturbed system as function of $k_z$. The difference between the perturbed and unperturbed band structures is shown in Fig. \ref{FIGS2}(b). The spin-orbit effect is now pronounced in the perturbed system (black, dashed lines), which allows us to fit $\alpha$ with much improved certainty. Note that the perturbation $U\Xi U^\dagger$ is very small as it only involves shifting the sub-bands within a single sub-band pair by at most a few meV. Moreover we have checked that the wavefunctions at finite $k_z$ are not significantly affected by the perturbation. Therefore, we are confident that this procedure allows for an accurate extraction of $\alpha$.

It can be shown that 
\begin{equation}
E_{ave}(k_z)= \frac{E_+ + E_-}{2} = E_o + \frac{\hbar^2 k_z^2}{2\overline{m}^*},    
\end{equation}
where $\overline{m}^*$ is the harmonic mean of the effective masses defined by
\begin{equation}
\frac{1}{\overline{m}^*} = \frac{1}{2}\left( \frac{1}{m_1^*} + \frac{1}{m_2^*} \right),    
\end{equation}
where $m_1^*$ and $m_2^*$ are the effective masses of the two sub-bands within a sub-band pair, respectively. Note that this result is independent of the spin-orbit coefficient, $\alpha$. We therefore choose to extract the harmonic mean effective mass, $\overline{m}^*$, of each sub-band pair, as oppose to each sub-bands effective mass since the result is independent of our fitting of the spin-orbit coefficient, $\alpha$. Note that we extract $\overline{m}^*$ from the original Hamiltonian's band structure and not the band structure of the perturbed Hamiltonian $\widetilde{H}$.

\section{Effective Mass}

\begin{figure}[t]
\begin{center}
\includegraphics[width=0.48\textwidth]{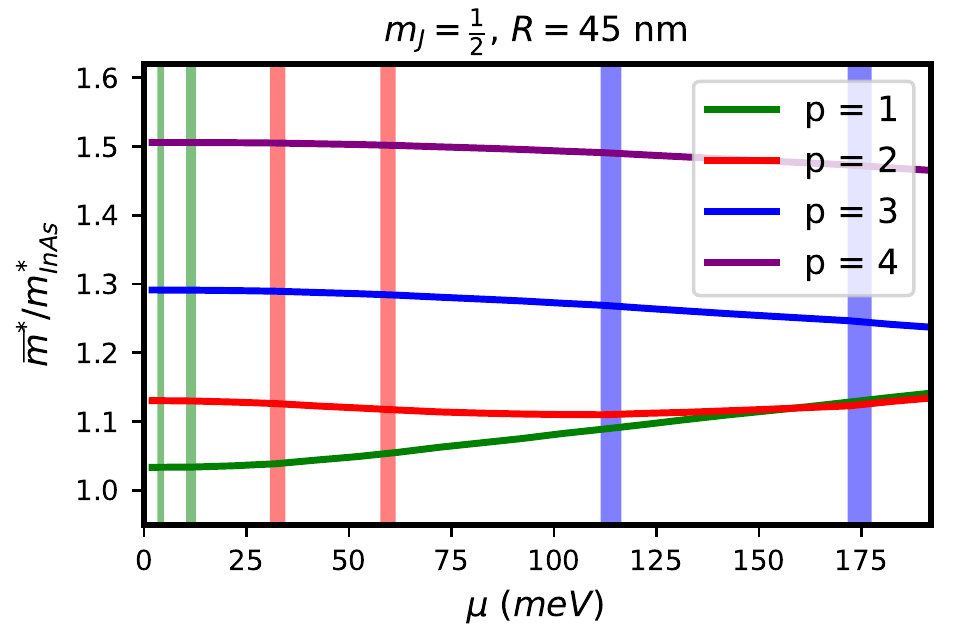}
\end{center}
\vspace{-4mm}
\caption{Hamonic mean effective mass of the first four $m_J = \frac{1}{2}$ sub-band pairs for a wire of radius $R = 45~$nm. The effective mass of the InAs conduction band edge is $m_{InAs}^* = 0.0229 ~ m_0$. The shaded regions correspond to the sub-band bottoms being within $\pm0.5~$meV of the chemical potential.}
\label{FIGS3}
\vspace{-1mm}
\end{figure}

A by-product of our parameter extraction procedure is the sub-band-dependent effective mass. In Fig \ref{FIGS3} we show the harmonic mean effective mass, $\overline{m}^*$,  for the first four $m_J=\frac{1}{2}$ sub-band pairs for a wire of radius $R=45~$nm. Note that the two sub-bands of each pair have slightly different effective masses, but only the harmonic mean is shown. The windows corresponding to the band bottom crossing the chemical potential are shaded (using matching colors). Note that the effective mass of the first two pairs is slightly larger than the bulk effective mass of the conduction band, while for $p=4$ it is already $50\%$ larger. In essence, we find the effective mass of the top occupied band increases with $\mu$ (i.e. with the work function difference). Note that in gated nanostructures the effective mass may also depend on the applied voltages.


\begin{thebibliography}{40}%
\makeatletter
\providecommand \@ifxundefined [1]{%
 \@ifx{#1\undefined}
}%
\providecommand \@ifnum [1]{%
 \ifnum #1\expandafter \@firstoftwo
 \else \expandafter \@secondoftwo
 \fi
}%
\providecommand \@ifx [1]{%
 \ifx #1\expandafter \@firstoftwo
 \else \expandafter \@secondoftwo
 \fi
}%
\providecommand \natexlab [1]{#1}%
\providecommand \enquote  [1]{``#1''}%
\providecommand \bibnamefont  [1]{#1}%
\providecommand \bibfnamefont [1]{#1}%
\providecommand \citenamefont [1]{#1}%
\providecommand \href@noop [0]{\@secondoftwo}%
\providecommand \href [0]{\begingroup \@sanitize@url \@href}%
\providecommand \@href[1]{\@@startlink{#1}\@@href}%
\providecommand \@@href[1]{\endgroup#1\@@endlink}%
\providecommand \@sanitize@url [0]{\catcode `\\12\catcode `\$12\catcode
  `\&12\catcode `\#12\catcode `\^12\catcode `\_12\catcode `\%12\relax}%
\providecommand \@@startlink[1]{}%
\providecommand \@@endlink[0]{}%
\providecommand \url  [0]{\begingroup\@sanitize@url \@url }%
\providecommand \@url [1]{\endgroup\@href {#1}{\urlprefix }}%
\providecommand \urlprefix  [0]{URL }%
\providecommand \Eprint [0]{\href }%
\providecommand \doibase [0]{http://dx.doi.org/}%
\providecommand \selectlanguage [0]{\@gobble}%
\providecommand \bibinfo  [0]{\@secondoftwo}%
\providecommand \bibfield  [0]{\@secondoftwo}%
\providecommand \translation [1]{[#1]}%
\providecommand \BibitemOpen [0]{}%
\providecommand \bibitemStop [0]{}%
\providecommand \bibitemNoStop [0]{.\EOS\space}%
\providecommand \EOS [0]{\spacefactor3000\relax}%
\providecommand \BibitemShut  [1]{\csname bibitem#1\endcsname}%
\let\auto@bib@innerbib\@empty
\bibitem [{\citenamefont {Read}\ and\ \citenamefont {Green}(2000)}]{Read2000}%
  \BibitemOpen
  \bibfield  {author} {\bibinfo {author} {\bibfnamefont {N.}~\bibnamefont
  {Read}}\ and\ \bibinfo {author} {\bibfnamefont {D.}~\bibnamefont {Green}},\
  }\href {\doibase 10.1103/PhysRevB.61.10267} {\bibfield  {journal} {\bibinfo
  {journal} {Phys. Rev. B}\ }\textbf {\bibinfo {volume} {61}},\ \bibinfo
  {pages} {10267} (\bibinfo {year} {2000})}\BibitemShut {NoStop}%
\bibitem [{\citenamefont {Kitaev}(2001)}]{Kitaev2001}%
  \BibitemOpen
  \bibfield  {author} {\bibinfo {author} {\bibfnamefont {A.~Y.}\ \bibnamefont
  {Kitaev}},\ }\href {\doibase 10.1070/1063-7869/44/10S/S29} {\bibfield
  {journal} {\bibinfo  {journal} {Physics-Uspekhi}\ }\textbf {\bibinfo {volume}
  {44}},\ \bibinfo {pages} {131} (\bibinfo {year} {2001})}\BibitemShut
  {NoStop}%
\bibitem [{\citenamefont {Nayak}\ \emph {et~al.}(2008)\citenamefont {Nayak},
  \citenamefont {Simon}, \citenamefont {Stern}, \citenamefont {Freedman},\ and\
  \citenamefont {Das~Sarma}}]{Nayak2008}%
  \BibitemOpen
  \bibfield  {author} {\bibinfo {author} {\bibfnamefont {C.}~\bibnamefont
  {Nayak}}, \bibinfo {author} {\bibfnamefont {S.~H.}\ \bibnamefont {Simon}},
  \bibinfo {author} {\bibfnamefont {A.}~\bibnamefont {Stern}}, \bibinfo
  {author} {\bibfnamefont {M.}~\bibnamefont {Freedman}}, \ and\ \bibinfo
  {author} {\bibfnamefont {S.}~\bibnamefont {Das~Sarma}},\ }\href {\doibase
  10.1103/RevModPhys.80.1083} {\bibfield  {journal} {\bibinfo  {journal} {Rev.
  Mod. Phys.}\ }\textbf {\bibinfo {volume} {80}},\ \bibinfo {pages} {1083}
  (\bibinfo {year} {2008})}\BibitemShut {NoStop}%
\bibitem [{\citenamefont {Das~Sarma}\ \emph {et~al.}(2015)\citenamefont
  {Das~Sarma}, \citenamefont {Freedman},\ and\ \citenamefont
  {Nayak}}]{DSarma2015}%
  \BibitemOpen
  \bibfield  {author} {\bibinfo {author} {\bibfnamefont {S.}~\bibnamefont
  {Das~Sarma}}, \bibinfo {author} {\bibfnamefont {M.}~\bibnamefont {Freedman}},
  \ and\ \bibinfo {author} {\bibfnamefont {C.}~\bibnamefont {Nayak}},\
  }\href@noop {} {\bibfield  {journal} {\bibinfo  {journal} {Npj Quantum
  Information}\ }\textbf {\bibinfo {volume} {1}},\ \bibinfo {pages} {15001}
  (\bibinfo {year} {2015})}\BibitemShut {NoStop}%
\bibitem [{\citenamefont {Sau}\ \emph {et~al.}(2010)\citenamefont {Sau},
  \citenamefont {Lutchyn}, \citenamefont {Tewari},\ and\ \citenamefont
  {Das~Sarma}}]{Sau2010a}%
  \BibitemOpen
  \bibfield  {author} {\bibinfo {author} {\bibfnamefont {J.~D.}\ \bibnamefont
  {Sau}}, \bibinfo {author} {\bibfnamefont {R.~M.}\ \bibnamefont {Lutchyn}},
  \bibinfo {author} {\bibfnamefont {S.}~\bibnamefont {Tewari}}, \ and\ \bibinfo
  {author} {\bibfnamefont {S.}~\bibnamefont {Das~Sarma}},\ }\href {\doibase
  10.1103/PhysRevLett.104.040502} {\bibfield  {journal} {\bibinfo  {journal}
  {Phys. Rev. Lett.}\ }\textbf {\bibinfo {volume} {104}},\ \bibinfo {pages}
  {040502} (\bibinfo {year} {2010})}\BibitemShut {NoStop}%
\bibitem [{\citenamefont {Lutchyn}\ \emph {et~al.}(2010)\citenamefont
  {Lutchyn}, \citenamefont {Sau},\ and\ \citenamefont
  {Das~Sarma}}]{Lutchyn2010}%
  \BibitemOpen
  \bibfield  {author} {\bibinfo {author} {\bibfnamefont {R.~M.}\ \bibnamefont
  {Lutchyn}}, \bibinfo {author} {\bibfnamefont {J.~D.}\ \bibnamefont {Sau}}, \
  and\ \bibinfo {author} {\bibfnamefont {S.}~\bibnamefont {Das~Sarma}},\ }\href
  {\doibase 10.1103/PhysRevLett.105.077001} {\bibfield  {journal} {\bibinfo
  {journal} {Phys. Rev. Lett.}\ }\textbf {\bibinfo {volume} {105}},\ \bibinfo
  {pages} {077001} (\bibinfo {year} {2010})}\BibitemShut {NoStop}%
\bibitem [{\citenamefont {Oreg}\ \emph {et~al.}(2010)\citenamefont {Oreg},
  \citenamefont {Refael},\ and\ \citenamefont {von Oppen}}]{Oreg2010}%
  \BibitemOpen
  \bibfield  {author} {\bibinfo {author} {\bibfnamefont {Y.}~\bibnamefont
  {Oreg}}, \bibinfo {author} {\bibfnamefont {G.}~\bibnamefont {Refael}}, \ and\
  \bibinfo {author} {\bibfnamefont {F.}~\bibnamefont {von Oppen}},\ }\href
  {\doibase 10.1103/PhysRevLett.105.177002} {\bibfield  {journal} {\bibinfo
  {journal} {Phys. Rev. Lett.}\ }\textbf {\bibinfo {volume} {105}},\ \bibinfo
  {pages} {177002} (\bibinfo {year} {2010})}\BibitemShut {NoStop}%
\bibitem [{\citenamefont {Mourik}\ \emph {et~al.}(2012)\citenamefont {Mourik},
  \citenamefont {Zuo}, \citenamefont {Frolov}, \citenamefont {Plissard},
  \citenamefont {Bakkers},\ and\ \citenamefont {Kouwenhoven}}]{Mourik2012}%
  \BibitemOpen
  \bibfield  {author} {\bibinfo {author} {\bibfnamefont {V.}~\bibnamefont
  {Mourik}}, \bibinfo {author} {\bibfnamefont {K.}~\bibnamefont {Zuo}},
  \bibinfo {author} {\bibfnamefont {S.~M.}\ \bibnamefont {Frolov}}, \bibinfo
  {author} {\bibfnamefont {S.~R.}\ \bibnamefont {Plissard}}, \bibinfo {author}
  {\bibfnamefont {E.~P. A.~M.}\ \bibnamefont {Bakkers}}, \ and\ \bibinfo
  {author} {\bibfnamefont {L.~P.}\ \bibnamefont {Kouwenhoven}},\ }\href
  {\doibase 10.1126/science.1222360} {\bibfield  {journal} {\bibinfo  {journal}
  {Science}\ }\textbf {\bibinfo {volume} {336}},\ \bibinfo {pages} {1003}
  (\bibinfo {year} {2012})}\BibitemShut {NoStop}%
\bibitem [{\citenamefont {Deng}\ \emph {et~al.}(2012)\citenamefont {Deng},
  \citenamefont {Yu}, \citenamefont {Huang}, \citenamefont {Larsson},
  \citenamefont {Caroff},\ and\ \citenamefont {Xu}}]{Deng2012}%
  \BibitemOpen
  \bibfield  {author} {\bibinfo {author} {\bibfnamefont {M.~T.}\ \bibnamefont
  {Deng}}, \bibinfo {author} {\bibfnamefont {C.~L.}\ \bibnamefont {Yu}},
  \bibinfo {author} {\bibfnamefont {G.~Y.}\ \bibnamefont {Huang}}, \bibinfo
  {author} {\bibfnamefont {M.}~\bibnamefont {Larsson}}, \bibinfo {author}
  {\bibfnamefont {P.}~\bibnamefont {Caroff}}, \ and\ \bibinfo {author}
  {\bibfnamefont {H.~Q.}\ \bibnamefont {Xu}},\ }\href {\doibase
  10.1021/nl303758w} {\bibfield  {journal} {\bibinfo  {journal} {Nano Letters}\
  }\textbf {\bibinfo {volume} {12}},\ \bibinfo {pages} {6414} (\bibinfo {year}
  {2012})}\BibitemShut {NoStop}%
\bibitem [{\citenamefont {Das}\ \emph {et~al.}(2012)\citenamefont {Das},
  \citenamefont {Ronen}, \citenamefont {Most}, \citenamefont {Oreg},
  \citenamefont {Heiblum},\ and\ \citenamefont {Shtrikman}}]{Das2012}%
  \BibitemOpen
  \bibfield  {author} {\bibinfo {author} {\bibfnamefont {A.}~\bibnamefont
  {Das}}, \bibinfo {author} {\bibfnamefont {Y.}~\bibnamefont {Ronen}}, \bibinfo
  {author} {\bibfnamefont {Y.}~\bibnamefont {Most}}, \bibinfo {author}
  {\bibfnamefont {Y.}~\bibnamefont {Oreg}}, \bibinfo {author} {\bibfnamefont
  {M.}~\bibnamefont {Heiblum}}, \ and\ \bibinfo {author} {\bibfnamefont
  {H.}~\bibnamefont {Shtrikman}},\ }\href@noop {} {\bibfield  {journal}
  {\bibinfo  {journal} {Nature Physics}\ }\textbf {\bibinfo {volume} {8}},\
  \bibinfo {pages} {887} (\bibinfo {year} {2012})}\BibitemShut {NoStop}%
\bibitem [{\citenamefont {Churchill}\ \emph {et~al.}(2013)\citenamefont
  {Churchill}, \citenamefont {Fatemi}, \citenamefont {Grove-Rasmussen},
  \citenamefont {Deng}, \citenamefont {Caroff}, \citenamefont {Xu},\ and\
  \citenamefont {Marcus}}]{Churchill2013}%
  \BibitemOpen
  \bibfield  {author} {\bibinfo {author} {\bibfnamefont {H.~O.~H.}\
  \bibnamefont {Churchill}}, \bibinfo {author} {\bibfnamefont {V.}~\bibnamefont
  {Fatemi}}, \bibinfo {author} {\bibfnamefont {K.}~\bibnamefont
  {Grove-Rasmussen}}, \bibinfo {author} {\bibfnamefont {M.~T.}\ \bibnamefont
  {Deng}}, \bibinfo {author} {\bibfnamefont {P.}~\bibnamefont {Caroff}},
  \bibinfo {author} {\bibfnamefont {H.~Q.}\ \bibnamefont {Xu}}, \ and\ \bibinfo
  {author} {\bibfnamefont {C.~M.}\ \bibnamefont {Marcus}},\ }\href {\doibase
  10.1103/PhysRevB.87.241401} {\bibfield  {journal} {\bibinfo  {journal} {Phys.
  Rev. B}\ }\textbf {\bibinfo {volume} {87}},\ \bibinfo {pages} {241401}
  (\bibinfo {year} {2013})}\BibitemShut {NoStop}%
\bibitem [{\citenamefont {Finck}\ \emph {et~al.}(2013)\citenamefont {Finck},
  \citenamefont {Van~Harlingen}, \citenamefont {Mohseni}, \citenamefont
  {Jung},\ and\ \citenamefont {Li}}]{Finck2013}%
  \BibitemOpen
  \bibfield  {author} {\bibinfo {author} {\bibfnamefont {A.~D.~K.}\
  \bibnamefont {Finck}}, \bibinfo {author} {\bibfnamefont {D.~J.}\ \bibnamefont
  {Van~Harlingen}}, \bibinfo {author} {\bibfnamefont {P.~K.}\ \bibnamefont
  {Mohseni}}, \bibinfo {author} {\bibfnamefont {K.}~\bibnamefont {Jung}}, \
  and\ \bibinfo {author} {\bibfnamefont {X.}~\bibnamefont {Li}},\ }\href
  {\doibase 10.1103/PhysRevLett.110.126406} {\bibfield  {journal} {\bibinfo
  {journal} {Phys. Rev. Lett.}\ }\textbf {\bibinfo {volume} {110}},\ \bibinfo
  {pages} {126406} (\bibinfo {year} {2013})}\BibitemShut {NoStop}%
\bibitem [{\citenamefont {Albrecht}\ \emph {et~al.}(2016)\citenamefont
  {Albrecht}, \citenamefont {Higginbotham}, \citenamefont {Madsen},
  \citenamefont {Kuemmeth}, \citenamefont {Jespersen}, \citenamefont {Nyg{\r
  a}rd}, \citenamefont {Krogstrup},\ and\ \citenamefont
  {Marcus}}]{Albrecht2016}%
  \BibitemOpen
  \bibfield  {author} {\bibinfo {author} {\bibfnamefont {S.~M.}\ \bibnamefont
  {Albrecht}}, \bibinfo {author} {\bibfnamefont {A.~P.}\ \bibnamefont
  {Higginbotham}}, \bibinfo {author} {\bibfnamefont {M.}~\bibnamefont
  {Madsen}}, \bibinfo {author} {\bibfnamefont {F.}~\bibnamefont {Kuemmeth}},
  \bibinfo {author} {\bibfnamefont {T.~S.}\ \bibnamefont {Jespersen}}, \bibinfo
  {author} {\bibfnamefont {J.}~\bibnamefont {Nyg{\r a}rd}}, \bibinfo {author}
  {\bibfnamefont {P.}~\bibnamefont {Krogstrup}}, \ and\ \bibinfo {author}
  {\bibfnamefont {C.~M.}\ \bibnamefont {Marcus}},\ }\href {\doibase
  10.1038/nature17162} {\bibfield  {journal} {\bibinfo  {journal} {Nature}\
  }\textbf {\bibinfo {volume} {531}},\ \bibinfo {pages} {206} (\bibinfo {year}
  {2016})}\BibitemShut {NoStop}%
\bibitem [{\citenamefont {Deng}\ \emph {et~al.}(2016)\citenamefont {Deng},
  \citenamefont {Vaitiekenas}, \citenamefont {Hansen}, \citenamefont {Danon},
  \citenamefont {Leijnse}, \citenamefont {Flensberg}, \citenamefont {Nyg{\r
  a}rd}, \citenamefont {Krogstrup},\ and\ \citenamefont {Marcus}}]{Deng2016}%
  \BibitemOpen
  \bibfield  {author} {\bibinfo {author} {\bibfnamefont {M.~T.}\ \bibnamefont
  {Deng}}, \bibinfo {author} {\bibfnamefont {S.}~\bibnamefont {Vaitiekenas}},
  \bibinfo {author} {\bibfnamefont {E.~B.}\ \bibnamefont {Hansen}}, \bibinfo
  {author} {\bibfnamefont {J.}~\bibnamefont {Danon}}, \bibinfo {author}
  {\bibfnamefont {M.}~\bibnamefont {Leijnse}}, \bibinfo {author} {\bibfnamefont
  {K.}~\bibnamefont {Flensberg}}, \bibinfo {author} {\bibfnamefont
  {J.}~\bibnamefont {Nyg{\r a}rd}}, \bibinfo {author} {\bibfnamefont
  {P.}~\bibnamefont {Krogstrup}}, \ and\ \bibinfo {author} {\bibfnamefont
  {C.~M.}\ \bibnamefont {Marcus}},\ }\href {\doibase 10.1126/science.aaf3961}
  {\bibfield  {journal} {\bibinfo  {journal} {Science}\ }\textbf {\bibinfo
  {volume} {354}},\ \bibinfo {pages} {1557} (\bibinfo {year}
  {2016})}\BibitemShut {NoStop}%
\bibitem [{\citenamefont {Nichele}\ \emph {et~al.}(2017)\citenamefont
  {Nichele}, \citenamefont {Drachmann}, \citenamefont {Whiticar}, \citenamefont
  {O'Farrell}, \citenamefont {Suominen}, \citenamefont {Fornieri},
  \citenamefont {Wang}, \citenamefont {Gardner}, \citenamefont {Thomas},
  \citenamefont {Hatke}, \citenamefont {Krogstrup}, \citenamefont {Manfra},
  \citenamefont {Flensberg},\ and\ \citenamefont {Marcus}}]{Nichele2017}%
  \BibitemOpen
  \bibfield  {author} {\bibinfo {author} {\bibfnamefont {F.}~\bibnamefont
  {Nichele}}, \bibinfo {author} {\bibfnamefont {A.~C.~C.}\ \bibnamefont
  {Drachmann}}, \bibinfo {author} {\bibfnamefont {A.~M.}\ \bibnamefont
  {Whiticar}}, \bibinfo {author} {\bibfnamefont {E.~C.~T.}\ \bibnamefont
  {O'Farrell}}, \bibinfo {author} {\bibfnamefont {H.~J.}\ \bibnamefont
  {Suominen}}, \bibinfo {author} {\bibfnamefont {A.}~\bibnamefont {Fornieri}},
  \bibinfo {author} {\bibfnamefont {T.}~\bibnamefont {Wang}}, \bibinfo {author}
  {\bibfnamefont {G.~C.}\ \bibnamefont {Gardner}}, \bibinfo {author}
  {\bibfnamefont {C.}~\bibnamefont {Thomas}}, \bibinfo {author} {\bibfnamefont
  {A.~T.}\ \bibnamefont {Hatke}}, \bibinfo {author} {\bibfnamefont
  {P.}~\bibnamefont {Krogstrup}}, \bibinfo {author} {\bibfnamefont {M.~J.}\
  \bibnamefont {Manfra}}, \bibinfo {author} {\bibfnamefont {K.}~\bibnamefont
  {Flensberg}}, \ and\ \bibinfo {author} {\bibfnamefont {C.~M.}\ \bibnamefont
  {Marcus}},\ }\href {\doibase 10.1103/PhysRevLett.119.136803} {\bibfield
  {journal} {\bibinfo  {journal} {Phys. Rev. Lett.}\ }\textbf {\bibinfo
  {volume} {119}},\ \bibinfo {pages} {136803} (\bibinfo {year}
  {2017})}\BibitemShut {NoStop}%
\bibitem [{\citenamefont {Zhang}\ \emph {et~al.}(2017)\citenamefont {Zhang},
  \citenamefont {Gul}, \citenamefont {Conesa-Boj}, \citenamefont {Nowak},
  \citenamefont {Wimmer}, \citenamefont {Zuo}, \citenamefont {Mourik},
  \citenamefont {de~Vries}, \citenamefont {van Veen}, \citenamefont {de~Moor},
  \citenamefont {Bommer}, \citenamefont {van Woerkom}, \citenamefont {Car},
  \citenamefont {Plissard}, \citenamefont {Bakkers}, \citenamefont
  {Quintero-Perez}, \citenamefont {Cassidy}, \citenamefont {Koelling},
  \citenamefont {Goswami}, \citenamefont {Watanabe}, \citenamefont
  {Taniguchi},\ and\ \citenamefont {Kouwenhoven}}]{Zhang2017}%
  \BibitemOpen
  \bibfield  {author} {\bibinfo {author} {\bibfnamefont {H.}~\bibnamefont
  {Zhang}}, \bibinfo {author} {\bibfnamefont {O.}~\bibnamefont {Gul}}, \bibinfo
  {author} {\bibfnamefont {S.}~\bibnamefont {Conesa-Boj}}, \bibinfo {author}
  {\bibfnamefont {M.~P.}\ \bibnamefont {Nowak}}, \bibinfo {author}
  {\bibfnamefont {M.}~\bibnamefont {Wimmer}}, \bibinfo {author} {\bibfnamefont
  {K.}~\bibnamefont {Zuo}}, \bibinfo {author} {\bibfnamefont {V.}~\bibnamefont
  {Mourik}}, \bibinfo {author} {\bibfnamefont {F.~K.}\ \bibnamefont
  {de~Vries}}, \bibinfo {author} {\bibfnamefont {J.}~\bibnamefont {van Veen}},
  \bibinfo {author} {\bibfnamefont {M.~W.}\ \bibnamefont {de~Moor}}, \bibinfo
  {author} {\bibfnamefont {J.~D.}\ \bibnamefont {Bommer}}, \bibinfo {author}
  {\bibfnamefont {D.~J.}\ \bibnamefont {van Woerkom}}, \bibinfo {author}
  {\bibfnamefont {D.}~\bibnamefont {Car}}, \bibinfo {author} {\bibfnamefont
  {S.~R.}\ \bibnamefont {Plissard}}, \bibinfo {author} {\bibfnamefont {E.~P.}\
  \bibnamefont {Bakkers}}, \bibinfo {author} {\bibfnamefont {M.}~\bibnamefont
  {Quintero-Perez}}, \bibinfo {author} {\bibfnamefont {M.~C.}\ \bibnamefont
  {Cassidy}}, \bibinfo {author} {\bibfnamefont {S.}~\bibnamefont {Koelling}},
  \bibinfo {author} {\bibfnamefont {S.}~\bibnamefont {Goswami}}, \bibinfo
  {author} {\bibfnamefont {K.}~\bibnamefont {Watanabe}}, \bibinfo {author}
  {\bibfnamefont {T.}~\bibnamefont {Taniguchi}}, \ and\ \bibinfo {author}
  {\bibfnamefont {L.~P.}\ \bibnamefont {Kouwenhoven}},\ }\href {\doibase
  10.1038/ncomms16025} {\bibfield  {journal} {\bibinfo  {journal} {Nat.
  Commun.}\ }\textbf {\bibinfo {volume} {8}},\ \bibinfo {pages} {16025}
  (\bibinfo {year} {2017})}\BibitemShut {NoStop}%
\bibitem [{\citenamefont {G\"{u}l}\ \emph {et~al.}(2018)\citenamefont
  {G\"{u}l}, \citenamefont {Zhang}, \citenamefont {Bommer}, \citenamefont
  {de~Moor}, \citenamefont {Car}, \citenamefont {Plissard}, \citenamefont
  {Bakkers}, \citenamefont {Geresdi}, \citenamefont {Watanabe}, \citenamefont
  {Taniguchi},\ and\ \citenamefont {Kouwenhoven}}]{Gul2018}%
  \BibitemOpen
  \bibfield  {author} {\bibinfo {author} {\bibfnamefont {O.}~\bibnamefont
  {G\"{u}l}}, \bibinfo {author} {\bibfnamefont {H.}~\bibnamefont {Zhang}},
  \bibinfo {author} {\bibfnamefont {J.~D.~S.}\ \bibnamefont {Bommer}}, \bibinfo
  {author} {\bibfnamefont {M.~W.~A.}\ \bibnamefont {de~Moor}}, \bibinfo
  {author} {\bibfnamefont {D.}~\bibnamefont {Car}}, \bibinfo {author}
  {\bibfnamefont {S.~R.}\ \bibnamefont {Plissard}}, \bibinfo {author}
  {\bibfnamefont {E.~P. A.~M.}\ \bibnamefont {Bakkers}}, \bibinfo {author}
  {\bibfnamefont {A.}~\bibnamefont {Geresdi}}, \bibinfo {author} {\bibfnamefont
  {K.}~\bibnamefont {Watanabe}}, \bibinfo {author} {\bibfnamefont
  {T.}~\bibnamefont {Taniguchi}}, \ and\ \bibinfo {author} {\bibfnamefont
  {L.~P.}\ \bibnamefont {Kouwenhoven}},\ }\href@noop {} {\bibfield  {journal}
  {\bibinfo  {journal} {Nature Nanotechnology}\ }\textbf {\bibinfo {volume}
  {13}},\ \bibinfo {pages} {192} (\bibinfo {year} {2018})}\BibitemShut
  {NoStop}%
\bibitem [{\citenamefont {Das~Sarma}\ \emph {et~al.}(2012)\citenamefont
  {Das~Sarma}, \citenamefont {Sau},\ and\ \citenamefont
  {Stanescu}}]{DSarma2012}%
  \BibitemOpen
  \bibfield  {author} {\bibinfo {author} {\bibfnamefont {S.}~\bibnamefont
  {Das~Sarma}}, \bibinfo {author} {\bibfnamefont {J.~D.}\ \bibnamefont {Sau}},
  \ and\ \bibinfo {author} {\bibfnamefont {T.~D.}\ \bibnamefont {Stanescu}},\
  }\href {\doibase 10.1103/PhysRevB.86.220506} {\bibfield  {journal} {\bibinfo
  {journal} {Phys. Rev. B}\ }\textbf {\bibinfo {volume} {86}},\ \bibinfo
  {pages} {220506} (\bibinfo {year} {2012})}\BibitemShut {NoStop}%
\bibitem [{\citenamefont {Kells}\ \emph {et~al.}(2012)\citenamefont {Kells},
  \citenamefont {Meidan},\ and\ \citenamefont {Brouwer}}]{Kells2012}%
  \BibitemOpen
  \bibfield  {author} {\bibinfo {author} {\bibfnamefont {G.}~\bibnamefont
  {Kells}}, \bibinfo {author} {\bibfnamefont {D.}~\bibnamefont {Meidan}}, \
  and\ \bibinfo {author} {\bibfnamefont {P.~W.}\ \bibnamefont {Brouwer}},\
  }\href {\doibase 10.1103/PhysRevB.86.100503} {\bibfield  {journal} {\bibinfo
  {journal} {Phys. Rev. B}\ }\textbf {\bibinfo {volume} {86}},\ \bibinfo
  {pages} {100503} (\bibinfo {year} {2012})}\BibitemShut {NoStop}%
\bibitem [{\citenamefont {Moore}\ \emph {et~al.}(2018)\citenamefont {Moore},
  \citenamefont {Stanescu},\ and\ \citenamefont {Tewari}}]{Moore2018}%
  \BibitemOpen
  \bibfield  {author} {\bibinfo {author} {\bibfnamefont {C.}~\bibnamefont
  {Moore}}, \bibinfo {author} {\bibfnamefont {T.~D.}\ \bibnamefont {Stanescu}},
  \ and\ \bibinfo {author} {\bibfnamefont {S.}~\bibnamefont {Tewari}},\ }\href
  {\doibase 10.1103/PhysRevB.97.165302} {\bibfield  {journal} {\bibinfo
  {journal} {Phys. Rev. B}\ }\textbf {\bibinfo {volume} {97}},\ \bibinfo
  {pages} {165302} (\bibinfo {year} {2018})}\BibitemShut {NoStop}%
\bibitem [{\citenamefont {Liu}\ \emph {et~al.}(2017)\citenamefont {Liu},
  \citenamefont {Sau}, \citenamefont {Stanescu},\ and\ \citenamefont
  {Das~Sarma}}]{Liu2017a}%
  \BibitemOpen
  \bibfield  {author} {\bibinfo {author} {\bibfnamefont {C.-X.}\ \bibnamefont
  {Liu}}, \bibinfo {author} {\bibfnamefont {J.~D.}\ \bibnamefont {Sau}},
  \bibinfo {author} {\bibfnamefont {T.~D.}\ \bibnamefont {Stanescu}}, \ and\
  \bibinfo {author} {\bibfnamefont {S.}~\bibnamefont {Das~Sarma}},\ }\href
  {\doibase 10.1103/PhysRevB.96.075161} {\bibfield  {journal} {\bibinfo
  {journal} {Phys. Rev. B}\ }\textbf {\bibinfo {volume} {96}},\ \bibinfo
  {pages} {075161} (\bibinfo {year} {2017})}\BibitemShut {NoStop}%
\bibitem [{\citenamefont {Vuik}\ \emph {et~al.}(2018)\citenamefont {Vuik},
  \citenamefont {Nijholt}, \citenamefont {Akhmerov},\ and\ \citenamefont
  {Wimmer}}]{Vuik2019}%
  \BibitemOpen
  \bibfield  {author} {\bibinfo {author} {\bibfnamefont {A.}~\bibnamefont
  {Vuik}}, \bibinfo {author} {\bibfnamefont {B.}~\bibnamefont {Nijholt}},
  \bibinfo {author} {\bibfnamefont {A.~R.}\ \bibnamefont {Akhmerov}}, \ and\
  \bibinfo {author} {\bibfnamefont {M.}~\bibnamefont {Wimmer}},\ }\href
  {https://arxiv.org/abs/1806.02801} {\bibfield  {journal} {\bibinfo  {journal}
  {arXiv:1806.02801}\ } (\bibinfo {year} {2018})}\BibitemShut {NoStop}%
\bibitem [{\citenamefont {Reeg}\ \emph
  {et~al.}(2018{\natexlab{a}})\citenamefont {Reeg}, \citenamefont {Dmytruk},
  \citenamefont {Chevallier}, \citenamefont {Loss},\ and\ \citenamefont
  {Klinovaja}}]{Reeg2018b}%
  \BibitemOpen
  \bibfield  {author} {\bibinfo {author} {\bibfnamefont {C.}~\bibnamefont
  {Reeg}}, \bibinfo {author} {\bibfnamefont {O.}~\bibnamefont {Dmytruk}},
  \bibinfo {author} {\bibfnamefont {D.}~\bibnamefont {Chevallier}}, \bibinfo
  {author} {\bibfnamefont {D.}~\bibnamefont {Loss}}, \ and\ \bibinfo {author}
  {\bibfnamefont {J.}~\bibnamefont {Klinovaja}},\ }\href {\doibase
  10.1103/PhysRevB.98.245407} {\bibfield  {journal} {\bibinfo  {journal} {Phys.
  Rev. B}\ }\textbf {\bibinfo {volume} {98}},\ \bibinfo {pages} {245407}
  (\bibinfo {year} {2018}{\natexlab{a}})}\BibitemShut {NoStop}%
\bibitem [{\citenamefont {Stanescu}\ and\ \citenamefont
  {Tewari}(2018)}]{Stanescu2018b}%
  \BibitemOpen
  \bibfield  {author} {\bibinfo {author} {\bibfnamefont {T.~D.}\ \bibnamefont
  {Stanescu}}\ and\ \bibinfo {author} {\bibfnamefont {S.}~\bibnamefont
  {Tewari}},\ }\href {https://arxiv.org/abs/1811.02557} {\bibfield  {journal}
  {\bibinfo  {journal} {arXiv:1811.02557}\ } (\bibinfo {year}
  {2018})}\BibitemShut {NoStop}%
\bibitem [{\citenamefont {Vaitiekenas}\ \emph {et~al.}(2018)\citenamefont
  {Vaitiekenas}, \citenamefont {Deng}, \citenamefont {Krogstrup},\ and\
  \citenamefont {Marcus}}]{Vaitiekenas2018}%
  \BibitemOpen
  \bibfield  {author} {\bibinfo {author} {\bibfnamefont {S.}~\bibnamefont
  {Vaitiekenas}}, \bibinfo {author} {\bibfnamefont {M.-T.}\ \bibnamefont
  {Deng}}, \bibinfo {author} {\bibfnamefont {P.}~\bibnamefont {Krogstrup}}, \
  and\ \bibinfo {author} {\bibfnamefont {C.~M.}\ \bibnamefont {Marcus}},\
  }\href {https://arxiv.org/abs/1809.05513} {\bibfield  {journal} {\bibinfo
  {journal} {arXiv:1809.05513}\ } (\bibinfo {year} {2018})}\BibitemShut
  {NoStop}%
\bibitem [{\citenamefont {Lutchyn}\ \emph {et~al.}(2018)\citenamefont
  {Lutchyn}, \citenamefont {Winkler}, \citenamefont {Heck}, \citenamefont
  {Karzig}, \citenamefont {Flensberg}, \citenamefont {Glazman},\ and\
  \citenamefont {Nayak}}]{Lutchyn2018b}%
  \BibitemOpen
  \bibfield  {author} {\bibinfo {author} {\bibfnamefont {R.~M.}\ \bibnamefont
  {Lutchyn}}, \bibinfo {author} {\bibfnamefont {G.~W.}\ \bibnamefont
  {Winkler}}, \bibinfo {author} {\bibfnamefont {B.~V.}\ \bibnamefont {Heck}},
  \bibinfo {author} {\bibfnamefont {T.}~\bibnamefont {Karzig}}, \bibinfo
  {author} {\bibfnamefont {K.}~\bibnamefont {Flensberg}}, \bibinfo {author}
  {\bibfnamefont {L.~I.}\ \bibnamefont {Glazman}}, \ and\ \bibinfo {author}
  {\bibfnamefont {C.}~\bibnamefont {Nayak}},\ }\href
  {https://arxiv.org/abs/1809.05512} {\bibfield  {journal} {\bibinfo  {journal}
  {arXiv:1809.05512}\ } (\bibinfo {year} {2018})}\BibitemShut {NoStop}%
\bibitem [{\citenamefont {Vuik}\ \emph {et~al.}(2016)\citenamefont {Vuik},
  \citenamefont {Eeltink}, \citenamefont {Akhmerov},\ and\ \citenamefont
  {Wimmer}}]{Vuik2016}%
  \BibitemOpen
  \bibfield  {author} {\bibinfo {author} {\bibfnamefont {A.}~\bibnamefont
  {Vuik}}, \bibinfo {author} {\bibfnamefont {D.}~\bibnamefont {Eeltink}},
  \bibinfo {author} {\bibfnamefont {A.~R.}\ \bibnamefont {Akhmerov}}, \ and\
  \bibinfo {author} {\bibfnamefont {M.}~\bibnamefont {Wimmer}},\ }\href
  {http://stacks.iop.org/1367-2630/18/i=3/a=033013} {\bibfield  {journal}
  {\bibinfo  {journal} {New Journal of Physics}\ }\textbf {\bibinfo {volume}
  {18}},\ \bibinfo {pages} {033013} (\bibinfo {year} {2016})}\BibitemShut
  {NoStop}%
\bibitem [{\citenamefont {Woods}\ \emph {et~al.}(2018)\citenamefont {Woods},
  \citenamefont {Stanescu},\ and\ \citenamefont {Das~Sarma}}]{Woods2018}%
  \BibitemOpen
  \bibfield  {author} {\bibinfo {author} {\bibfnamefont {B.~D.}\ \bibnamefont
  {Woods}}, \bibinfo {author} {\bibfnamefont {T.~D.}\ \bibnamefont {Stanescu}},
  \ and\ \bibinfo {author} {\bibfnamefont {S.}~\bibnamefont {Das~Sarma}},\
  }\href {\doibase 10.1103/PhysRevB.98.035428} {\bibfield  {journal} {\bibinfo
  {journal} {Phys. Rev. B}\ }\textbf {\bibinfo {volume} {98}},\ \bibinfo
  {pages} {035428} (\bibinfo {year} {2018})}\BibitemShut {NoStop}%
\bibitem [{\citenamefont {Mikkelsen}\ \emph {et~al.}(2018)\citenamefont
  {Mikkelsen}, \citenamefont {Kotetes}, \citenamefont {Krogstrup},\ and\
  \citenamefont {Flensberg}}]{Mikkelsen2018}%
  \BibitemOpen
  \bibfield  {author} {\bibinfo {author} {\bibfnamefont {A.~E.~G.}\
  \bibnamefont {Mikkelsen}}, \bibinfo {author} {\bibfnamefont {P.}~\bibnamefont
  {Kotetes}}, \bibinfo {author} {\bibfnamefont {P.}~\bibnamefont {Krogstrup}},
  \ and\ \bibinfo {author} {\bibfnamefont {K.}~\bibnamefont {Flensberg}},\
  }\href {\doibase 10.1103/PhysRevX.8.031040} {\bibfield  {journal} {\bibinfo
  {journal} {Phys. Rev. X}\ }\textbf {\bibinfo {volume} {8}},\ \bibinfo {pages}
  {031040} (\bibinfo {year} {2018})}\BibitemShut {NoStop}%
\bibitem [{\citenamefont {Antipov}\ \emph {et~al.}(2018)\citenamefont
  {Antipov}, \citenamefont {Bargerbos}, \citenamefont {Winkler}, \citenamefont
  {Bauer}, \citenamefont {Rossi},\ and\ \citenamefont {Lutchyn}}]{Antipov2018}%
  \BibitemOpen
  \bibfield  {author} {\bibinfo {author} {\bibfnamefont {A.~E.}\ \bibnamefont
  {Antipov}}, \bibinfo {author} {\bibfnamefont {A.}~\bibnamefont {Bargerbos}},
  \bibinfo {author} {\bibfnamefont {G.~W.}\ \bibnamefont {Winkler}}, \bibinfo
  {author} {\bibfnamefont {B.}~\bibnamefont {Bauer}}, \bibinfo {author}
  {\bibfnamefont {E.}~\bibnamefont {Rossi}}, \ and\ \bibinfo {author}
  {\bibfnamefont {R.~M.}\ \bibnamefont {Lutchyn}},\ }\href {\doibase
  10.1103/PhysRevX.8.031041} {\bibfield  {journal} {\bibinfo  {journal} {Phys.
  Rev. X}\ }\textbf {\bibinfo {volume} {8}},\ \bibinfo {pages} {031041}
  (\bibinfo {year} {2018})}\BibitemShut {NoStop}%
\bibitem [{\citenamefont {Winkler}(2003)}]{Winkler2003}%
  \BibitemOpen
  \bibfield  {author} {\bibinfo {author} {\bibfnamefont {R.}~\bibnamefont
  {Winkler}},\ }\href@noop {} {\emph {\bibinfo {title} {Spin-Orbit Coupling
  Effects in Two-Dimensional Electron and Hole Systems}}}\ (\bibinfo
  {publisher} {Springer},\ \bibinfo {year} {2003})\BibitemShut {NoStop}%
\bibitem [{\citenamefont {Kohn}\ and\ \citenamefont
  {Luttinger}(1955)}]{Kohn1955}%
  \BibitemOpen
  \bibfield  {author} {\bibinfo {author} {\bibfnamefont {W.}~\bibnamefont
  {Kohn}}\ and\ \bibinfo {author} {\bibfnamefont {J.~M.}\ \bibnamefont
  {Luttinger}},\ }\href {\doibase 10.1103/PhysRev.98.915} {\bibfield  {journal}
  {\bibinfo  {journal} {Phys. Rev.}\ }\textbf {\bibinfo {volume} {98}},\
  \bibinfo {pages} {915} (\bibinfo {year} {1955})}\BibitemShut {NoStop}%
\bibitem [{\citenamefont {Luttinger}(1956)}]{Luttinger1956}%
  \BibitemOpen
  \bibfield  {author} {\bibinfo {author} {\bibfnamefont {J.~M.}\ \bibnamefont
  {Luttinger}},\ }\href {\doibase 10.1103/PhysRev.102.1030} {\bibfield
  {journal} {\bibinfo  {journal} {Phys. Rev.}\ }\textbf {\bibinfo {volume}
  {102}},\ \bibinfo {pages} {1030} (\bibinfo {year} {1956})}\BibitemShut
  {NoStop}%
\bibitem [{\citenamefont {Chang}(1988)}]{Chang1988}%
  \BibitemOpen
  \bibfield  {author} {\bibinfo {author} {\bibfnamefont {Y.-C.}\ \bibnamefont
  {Chang}},\ }\href {\doibase 10.1103/PhysRevB.37.8215} {\bibfield  {journal}
  {\bibinfo  {journal} {Phys. Rev. B}\ }\textbf {\bibinfo {volume} {37}},\
  \bibinfo {pages} {8215} (\bibinfo {year} {1988})}\BibitemShut {NoStop}%
\bibitem [{\citenamefont {Loehr}(1994)}]{Loehr1994}%
  \BibitemOpen
  \bibfield  {author} {\bibinfo {author} {\bibfnamefont {J.~P.}\ \bibnamefont
  {Loehr}},\ }\href {\doibase 10.1103/PhysRevB.50.5429} {\bibfield  {journal}
  {\bibinfo  {journal} {Phys. Rev. B}\ }\textbf {\bibinfo {volume} {50}},\
  \bibinfo {pages} {5429} (\bibinfo {year} {1994})}\BibitemShut {NoStop}%
\bibitem [{\citenamefont {Marquardt}\ \emph {et~al.}(2008)\citenamefont
  {Marquardt}, \citenamefont {Mourad}, \citenamefont {Schulz}, \citenamefont
  {Hickel}, \citenamefont {Czycholl},\ and\ \citenamefont
  {Neugebauer}}]{Marquardt2008}%
  \BibitemOpen
  \bibfield  {author} {\bibinfo {author} {\bibfnamefont {O.}~\bibnamefont
  {Marquardt}}, \bibinfo {author} {\bibfnamefont {D.}~\bibnamefont {Mourad}},
  \bibinfo {author} {\bibfnamefont {S.}~\bibnamefont {Schulz}}, \bibinfo
  {author} {\bibfnamefont {T.}~\bibnamefont {Hickel}}, \bibinfo {author}
  {\bibfnamefont {G.}~\bibnamefont {Czycholl}}, \ and\ \bibinfo {author}
  {\bibfnamefont {J.}~\bibnamefont {Neugebauer}},\ }\href {\doibase
  10.1103/PhysRevB.78.235302} {\bibfield  {journal} {\bibinfo  {journal} {Phys.
  Rev. B}\ }\textbf {\bibinfo {volume} {78}},\ \bibinfo {pages} {235302}
  (\bibinfo {year} {2008})}\BibitemShut {NoStop}%
\bibitem [{\citenamefont {Luo}\ \emph {et~al.}(2016)\citenamefont {Luo},
  \citenamefont {Liao},\ and\ \citenamefont {Xu}}]{Luo2016}%
  \BibitemOpen
  \bibfield  {author} {\bibinfo {author} {\bibfnamefont {N.}~\bibnamefont
  {Luo}}, \bibinfo {author} {\bibfnamefont {G.}~\bibnamefont {Liao}}, \ and\
  \bibinfo {author} {\bibfnamefont {H.~Q.}\ \bibnamefont {Xu}},\ }\href
  {\doibase 10.1063/1.4972987} {\bibfield  {journal} {\bibinfo  {journal} {AIP
  Advances}\ }\textbf {\bibinfo {volume} {6}},\ \bibinfo {pages} {125109}
  (\bibinfo {year} {2016})},\ \Eprint
  {http://arxiv.org/abs/https://doi.org/10.1063/1.4972987}
  {https://doi.org/10.1063/1.4972987} \BibitemShut {NoStop}%
\bibitem [{\citenamefont {Cole}\ \emph {et~al.}(2015)\citenamefont {Cole},
  \citenamefont {Das~Sarma},\ and\ \citenamefont {Stanescu}}]{Cole2015}%
  \BibitemOpen
  \bibfield  {author} {\bibinfo {author} {\bibfnamefont {W.~S.}\ \bibnamefont
  {Cole}}, \bibinfo {author} {\bibfnamefont {S.}~\bibnamefont {Das~Sarma}}, \
  and\ \bibinfo {author} {\bibfnamefont {T.~D.}\ \bibnamefont {Stanescu}},\
  }\href {\doibase 10.1103/PhysRevB.92.174511} {\bibfield  {journal} {\bibinfo
  {journal} {Phys. Rev. B}\ }\textbf {\bibinfo {volume} {92}},\ \bibinfo
  {pages} {174511} (\bibinfo {year} {2015})}\BibitemShut {NoStop}%
\bibitem [{\citenamefont {Stanescu}\ and\ \citenamefont
  {Das~Sarma}(2017)}]{Stanescu2017a}%
  \BibitemOpen
  \bibfield  {author} {\bibinfo {author} {\bibfnamefont {T.~D.}\ \bibnamefont
  {Stanescu}}\ and\ \bibinfo {author} {\bibfnamefont {S.}~\bibnamefont
  {Das~Sarma}},\ }\href {\doibase 10.1103/PhysRevB.96.014510} {\bibfield
  {journal} {\bibinfo  {journal} {Phys. Rev. B}\ }\textbf {\bibinfo {volume}
  {96}},\ \bibinfo {pages} {014510} (\bibinfo {year} {2017})}\BibitemShut
  {NoStop}%
\bibitem [{\citenamefont {Reeg}\ \emph
  {et~al.}(2018{\natexlab{b}})\citenamefont {Reeg}, \citenamefont {Loss},\ and\
  \citenamefont {Klinovaja}}]{Reeg2018a}%
  \BibitemOpen
  \bibfield  {author} {\bibinfo {author} {\bibfnamefont {C.}~\bibnamefont
  {Reeg}}, \bibinfo {author} {\bibfnamefont {D.}~\bibnamefont {Loss}}, \ and\
  \bibinfo {author} {\bibfnamefont {J.}~\bibnamefont {Klinovaja}},\ }\href
  {\doibase 10.1103/PhysRevB.97.165425} {\bibfield  {journal} {\bibinfo
  {journal} {Phys. Rev. B}\ }\textbf {\bibinfo {volume} {97}},\ \bibinfo
  {pages} {165425} (\bibinfo {year} {2018}{\natexlab{b}})}\BibitemShut
  {NoStop}%
\end{thebibliography}

\begin{thebibliography}{8}%
\makeatletter
\providecommand \@ifxundefined [1]{%
 \@ifx{#1\undefined}
}%
\providecommand \@ifnum [1]{%
 \ifnum #1\expandafter \@firstoftwo
 \else \expandafter \@secondoftwo
 \fi
}%
\providecommand \@ifx [1]{%
 \ifx #1\expandafter \@firstoftwo
 \else \expandafter \@secondoftwo
 \fi
}%
\providecommand \natexlab [1]{#1}%
\providecommand \enquote  [1]{``#1''}%
\providecommand \bibnamefont  [1]{#1}%
\providecommand \bibfnamefont [1]{#1}%
\providecommand \citenamefont [1]{#1}%
\providecommand \href@noop [0]{\@secondoftwo}%
\providecommand \href [0]{\begingroup \@sanitize@url \@href}%
\providecommand \@href[1]{\@@startlink{#1}\@@href}%
\providecommand \@@href[1]{\endgroup#1\@@endlink}%
\providecommand \@sanitize@url [0]{\catcode `\\12\catcode `\$12\catcode
  `\&12\catcode `\#12\catcode `\^12\catcode `\_12\catcode `\%12\relax}%
\providecommand \@@startlink[1]{}%
\providecommand \@@endlink[0]{}%
\providecommand \url  [0]{\begingroup\@sanitize@url \@url }%
\providecommand \@url [1]{\endgroup\@href {#1}{\urlprefix }}%
\providecommand \urlprefix  [0]{URL }%
\providecommand \Eprint [0]{\href }%
\providecommand \doibase [0]{http://dx.doi.org/}%
\providecommand \selectlanguage [0]{\@gobble}%
\providecommand \bibinfo  [0]{\@secondoftwo}%
\providecommand \bibfield  [0]{\@secondoftwo}%
\providecommand \translation [1]{[#1]}%
\providecommand \BibitemOpen [0]{}%
\providecommand \bibitemStop [0]{}%
\providecommand \bibitemNoStop [0]{.\EOS\space}%
\providecommand \EOS [0]{\spacefactor3000\relax}%
\providecommand \BibitemShut  [1]{\csname bibitem#1\endcsname}%
\let\auto@bib@innerbib\@empty
\bibitem [{\citenamefont {Kane}(1957)}]{Kane1957}%
  \BibitemOpen
  \bibfield  {author} {\bibinfo {author} {\bibfnamefont {E.~O.}\ \bibnamefont
  {Kane}},\ }\href {\doibase https://doi.org/10.1016/0022-3697(57)90013-6}
  {\bibfield  {journal} {\bibinfo  {journal} {Journal of Physics and Chemistry
  of Solids}\ }\textbf {\bibinfo {volume} {1}},\ \bibinfo {pages} {249 }
  (\bibinfo {year} {1957})}\BibitemShut {NoStop}%
\bibitem [{\citenamefont {Winkler}(2003)}]{Winkler2003A}%
  \BibitemOpen
  \bibfield  {author} {\bibinfo {author} {\bibfnamefont {R.}~\bibnamefont
  {Winkler}},\ }\href@noop {} {\emph {\bibinfo {title} {Spin-Orbit Coupling
  Effects in Two-Dimensional Electron and Hole Systems}}}\ (\bibinfo
  {publisher} {Springer},\ \bibinfo {year} {2003})\BibitemShut {NoStop}%
\bibitem [{\citenamefont {Luo}\ \emph {et~al.}(2016)\citenamefont {Luo},
  \citenamefont {Liao},\ and\ \citenamefont {Xu}}]{Luo2016A}%
  \BibitemOpen
  \bibfield  {author} {\bibinfo {author} {\bibfnamefont {N.}~\bibnamefont
  {Luo}}, \bibinfo {author} {\bibfnamefont {G.}~\bibnamefont {Liao}}, \ and\
  \bibinfo {author} {\bibfnamefont {H.~Q.}\ \bibnamefont {Xu}},\ }\href
  {\doibase 10.1063/1.4972987} {\bibfield  {journal} {\bibinfo  {journal} {AIP
  Advances}\ }\textbf {\bibinfo {volume} {6}},\ \bibinfo {pages} {125109}
  (\bibinfo {year} {2016})},\ \Eprint
  {http://arxiv.org/abs/https://doi.org/10.1063/1.4972987}
  {https://doi.org/10.1063/1.4972987} \BibitemShut {NoStop}%
\bibitem [{\citenamefont {Ivashev}(2016)}]{Ivashev2016}%
  \BibitemOpen
  \bibfield  {author} {\bibinfo {author} {\bibfnamefont {I.}~\bibnamefont
  {Ivashev}},\ }\emph {\bibinfo {title} {{Theoretical investigations of Zinc
  Blende and Wurtzite semiconductor quantum wells on the rotated
  substrates}}},\ \href {https://scholars.wlu.ca/etd/1833} {Master's thesis},\
  \bibinfo  {school} {Wilfrid Laurier University}, \bibinfo {address} {Canada}
  (\bibinfo {year} {2016})\BibitemShut {NoStop}%
\bibitem [{\citenamefont {Arsoski}\ \emph {et~al.}(2015)\citenamefont
  {Arsoski}, \citenamefont {Čukarić}, \citenamefont {Tadić},\ and\
  \citenamefont {Peeters}}]{Arsoski2015}%
  \BibitemOpen
  \bibfield  {author} {\bibinfo {author} {\bibfnamefont {V.}~\bibnamefont
  {Arsoski}}, \bibinfo {author} {\bibfnamefont {N.}~\bibnamefont {Čukarić}},
  \bibinfo {author} {\bibfnamefont {M.}~\bibnamefont {Tadić}}, \ and\ \bibinfo
  {author} {\bibfnamefont {F.}~\bibnamefont {Peeters}},\ }\href {\doibase
  https://doi.org/10.1016/j.cpc.2015.08.002} {\bibfield  {journal} {\bibinfo
  {journal} {Computer Physics Communications}\ }\textbf {\bibinfo {volume}
  {197}},\ \bibinfo {pages} {17 } (\bibinfo {year} {2015})}\BibitemShut
  {NoStop}%
\bibitem [{\citenamefont {Winkler}\ \emph {et~al.}(2017)\citenamefont
  {Winkler}, \citenamefont {Varjas}, \citenamefont {Skolasinski}, \citenamefont
  {Soluyanov}, \citenamefont {Troyer},\ and\ \citenamefont
  {Wimmer}}]{Winkler2017}%
  \BibitemOpen
  \bibfield  {author} {\bibinfo {author} {\bibfnamefont {G.~W.}\ \bibnamefont
  {Winkler}}, \bibinfo {author} {\bibfnamefont {D.}~\bibnamefont {Varjas}},
  \bibinfo {author} {\bibfnamefont {R.}~\bibnamefont {Skolasinski}}, \bibinfo
  {author} {\bibfnamefont {A.~A.}\ \bibnamefont {Soluyanov}}, \bibinfo {author}
  {\bibfnamefont {M.}~\bibnamefont {Troyer}}, \ and\ \bibinfo {author}
  {\bibfnamefont {M.}~\bibnamefont {Wimmer}},\ }\href {\doibase
  10.1103/PhysRevLett.119.037701} {\bibfield  {journal} {\bibinfo  {journal}
  {Phys. Rev. Lett.}\ }\textbf {\bibinfo {volume} {119}},\ \bibinfo {pages}
  {037701} (\bibinfo {year} {2017})}\BibitemShut {NoStop}%
\bibitem [{\citenamefont {Laliena}\ and\ \citenamefont
  {Campo}(2018)}]{Laliena2018}%
  \BibitemOpen
  \bibfield  {author} {\bibinfo {author} {\bibfnamefont {V.}~\bibnamefont
  {Laliena}}\ and\ \bibinfo {author} {\bibfnamefont {J.}~\bibnamefont
  {Campo}},\ }\href {http://stacks.iop.org/1751-8121/51/i=32/a=325203}
  {\bibfield  {journal} {\bibinfo  {journal} {Journal of Physics A:
  Mathematical and Theoretical}\ }\textbf {\bibinfo {volume} {51}},\ \bibinfo
  {pages} {325203} (\bibinfo {year} {2018})}\BibitemShut {NoStop}%
\bibitem [{\citenamefont {Lutchyn}\ \emph {et~al.}(2018)\citenamefont
  {Lutchyn}, \citenamefont {Winkler}, \citenamefont {Heck}, \citenamefont
  {Karzig}, \citenamefont {Flensberg}, \citenamefont {Glazman},\ and\
  \citenamefont {Nayak}}]{Lutchyn2018C}%
  \BibitemOpen
  \bibfield  {author} {\bibinfo {author} {\bibfnamefont {R.~M.}\ \bibnamefont
  {Lutchyn}}, \bibinfo {author} {\bibfnamefont {G.~W.}\ \bibnamefont
  {Winkler}}, \bibinfo {author} {\bibfnamefont {B.~V.}\ \bibnamefont {Heck}},
  \bibinfo {author} {\bibfnamefont {T.}~\bibnamefont {Karzig}}, \bibinfo
  {author} {\bibfnamefont {K.}~\bibnamefont {Flensberg}}, \bibinfo {author}
  {\bibfnamefont {L.~I.}\ \bibnamefont {Glazman}}, \ and\ \bibinfo {author}
  {\bibfnamefont {C.}~\bibnamefont {Nayak}},\ }\href
  {https://arxiv.org/abs/1809.05512} {\bibfield  {journal} {\bibinfo  {journal}
  {arXiv:1809.05512}\ } (\bibinfo {year} {2018})}\BibitemShut {NoStop}%
\end{thebibliography}
\end{document}